\documentclass{emulateapj}

\usepackage{xspace}
\usepackage{color}
\usepackage{rotating}
\usepackage{amsmath}
\usepackage{subfigure}
\PassOptionsToPackage{hyphens}{url}\usepackage{hyperref}

\slugcomment{Draft version \today}

\shorttitle{Exo-Aurorae on Proxima Centauri \MakeLowercase{b}}
\shortauthors{Luger et al.}

\begin{document}

%% Title, authors, affils
\title{The Pale Green Dot: A Method to Characterize Proxima Centauri \MakeLowercase{b} using Exo-Aurorae}

\author{Rodrigo Luger\altaffilmark{1,3,4,5}}
\author{Jacob Lustig-Yaeger\altaffilmark{1,3,4}}
\author{David P. Fleming\altaffilmark{1,3}}
\author{Matt A. Tilley\altaffilmark{2,3,4}}
\author{Eric Agol\altaffilmark{1,3,4}}
\author{Victoria S. Meadows\altaffilmark{1,3,4}}
\author{Russell Deitrick\altaffilmark{1,3,4}}
\author{Rory Barnes\altaffilmark{1,3,4}}

\altaffiltext{1}{Astronomy Department, University of Washington, Box 951580, Seattle, WA 98195}
\altaffiltext{2}{Department of Earth \& Space Sciences, University of Washington, Box 351310, Seattle, WA 98195}
\altaffiltext{3}{NASA Astrobiology Institute -- Virtual Planetary Laboratory Lead Team, USA}
\altaffiltext{4}{Astrobiology Program, University of Washington, 3910 15th Ave. NE, Box 351580, Seattle, WA 98195, USA}
\altaffiltext{5}{\email{rodluger@uw.edu}}

%% Abstract %%
\begin{abstract}
We examine the feasibility of detecting auroral emission from the potentially habitable exoplanet Proxima Centauri b. Detection of aurorae would yield an independent confirmation of the planet's existence, constrain the presence and composition of its atmosphere, and determine the planet's eccentricity and inclination, thereby breaking the mass-inclination degeneracy. If Proxima Centauri b is a terrestrial world with an Earth-like atmosphere and magnetic field, we estimate the power at the 5577\AA\ OI auroral line is on the order of 0.1~TW under steady-state stellar wind, or ${\sim} 100 {\times}$ stronger than that on Earth. This corresponds to a planet-star contrast ratio of $10^{-6}-10^{-7}$ in a narrow band about the 5577\AA\ line, although higher contrast ($10^{-4}-10^{-5}$) may be possible during periods of strong magnetospheric disturbance (auroral power $1-10$~TW). We searched the Proxima Centauri b HARPS data for the 5577\AA\ line and for other prominent oxygen and nitrogen lines, but find no signal, indicating that the OI auroral line contrast must be lower than $2\times 10^{-2}$ (with power $\lesssim$ 3,000~TW), consistent with our predictions. We find that observations of 0.1~TW auroral emission lines are likely infeasible with current and planned telescopes.  However, future observations with a space-based coronagraphic telescope or a ground-based extremely large telescope (ELT) with a coronagraph could push sensitivity down to terawatt oxygen aurorae (contrast $7\times 10^{-6}$) with exposure times of ${\sim} 1$ day. If a coronagraph design contrast of $10^{-7}$ can be achieved with negligible instrumental noise, a future concept ELT could observe steady-state auroral emission in a few nights.\\[0in]
\end{abstract}

%% Keywords %%

%% Introduction %%
\section{Introduction\label{sec:intro}}

The discovery of Proxima Centauri b (henceforth `Proxima Cen b'), only 1.3pc distant from the Sun \citep{Anglada-Escude2016}, ushers in a new era of characterization of nearby potentially habitable exoplanets. Although Proxima Cen b is not known to transit---making transmission spectroscopy impossible---it is an ideal candidate for high-contrast direct spectroscopy using an extremely large coronagraph-equipped telescope. However, even with the enhancement in angular resolution provided by the proximity of its host star, Proxima Cen b's close-in orbit \citep[$a = 0.0485$ AU;][]{Anglada-Escude2016} precludes imaging with current coronagraphs, such as the Gemini Planet Imager \citep[GPI;][]{Macintosh2014} and the Very Large Telescope's Spectro-Polarimetric High-contrast Exoplanet REsearch facility \citep[VLT-SPHERE;][]{Beuzit2008}, which operate primarily in the near-infrared. This is in part due to the poorer Strehl ratios currently achievable at visible wavelengths with ground-based adaptive optics (AO) systems\footnote{See, e.g., \url{https://www.eso.org/sci/facilities/paranal/instruments/sphere/overview.html}}. Consequently, in advance of larger diameter ground- and space-based telescopes, and improvements in visible AO systems, we must initially consider observations that do not rely on transits or current coronagraphy to search for and characterize the atmosphere of Proxima Cen b.

Phase curves may offer one of the first means to study the atmosphere of Proxima Cen b \citep{Turbet2016,Kreidberg2016,Meadows2016} by potentially showing the reduction in day-night thermal emission contrast associated with an atmosphere. Phase curves have proven to be a successful means to characterize the atmospheres of planets larger and hotter than Proxima Cen b \citep{Cowan2007, Knutson2007, Knutson2008, Crossfield2010, Brogi2012, Zellem2014, Stevenson2014}, including ones that do not transit \citep{Selsis2011, Faigler2011, Maurin2012, Brogi2014}. However, the expected planet-star contrast ratio in the visible and NIR due to reflected stellar radiation is likely to be below the anticipated systematic noise floor for JWST/NIRSpec \citep{Meadows2016}, and although the planet-star contrast ratio becomes quite favorable beyond $10\mu$m, where the planetary thermal emission peaks, mid-IR phase curves will require JWST/MIRI.  

To complement the anticipated JWST thermal phase curve measurements, in this work we explore the possibility of directly detecting optical auroral emission from the atmosphere of Proxima Cen b using high-resolution optical spectroscopy. Numerous studies have investigated exoplanet aurorae in the radio due to cyclotron and synchrotron emission to constrain the planetary magnetic field \citep[e.g.][]{Bastian2000,Grie2007,Zarka2007,Hess2011,Driscoll2011,Griessmeier2015}. Others have explored the detectability of optical and UV auroral emission from hot Jupiters, including \citet{France2010}, who searched for far-UV auroral and dayglow H$_2$ emission from the hot Jupiter HD 209458b and placed upper limits on its magnetic field strength, and \citet{Menager2013}, who studied the detectability of Lyman $\alpha$ auroral emission from HD 209458b and HD 189733b.
Some studies have also investigated auroral emission from terrestrial planets, including \citet{Smith2004}, who modeled the role of aurorae in redistributing high energy incident stellar flux to the surface of rocky exoplanets, and \citet{Bernard2014}, who investigated how the detection of the green oxygen airglow line could be used to infer the presence of planetary hydrogen coronae on CO$_2$-dominated planets. Finally, \citet{SparksFord2002} suggested that exoplanet airglow and/or aurorae could be detected with a combination of high contrast imaging and high dispersion spectroscopy. However, a detailed calculation of the expected auroral signal strength on a nearby terrestrial exoplanet and the feasibility of its detection has not yet been fully performed.

\begin{deluxetable}{lcr}
\tablewidth{\linewidth}
\tablecaption{Proxima Centauri b properties}
\tablenum{1}
\tablehead{\colhead{Property} & \colhead{Value$^{\dagger}$} & \colhead{1$\sigma$ Interval}} 
\startdata
Distance from Earth (pc)        & 1.295     &                   \\
Host spectral type              & M5.5V     &                   \\
Host mass, $M_\star$ (M$_\odot$)& 0.120     & [0.105 -- 0.135]  \\
Period, $P$ (days)              & 11.186    & [11.184 -- 11.187]\\
Semi-major axis, $a$ (AU)       & 0.0485    & [0.0434 -- 0.0526]\\
Minimum mass, $m_p\sin{i}$ (M$_\oplus$)& 1.27 & [1.10 -- 1.46]  \\
Radius, $R_p$ (R$_\oplus$)      & Unknown   & [0.94 -- 1.40]$^{\ddagger}$    \\
Eccentricity, $e$               & $<0.35$   &                   \\
Mean longitude, $\lambda$ ($^\circ$)& 110   & [102 -- 118]      \\
Inclination, $i$ ($^\circ$)     & Unknown   & [0 -- 90]
\enddata
\tablenotetext{$\dagger$}{Values from \citet{Anglada-Escude2016} unless otherwise noted.}
\tablenotetext{$\ddagger$}{Plausible range from \citet{Brugger2016}, assuming $m_p = 1.27\mathrm{M}_\oplus$.}
\label{tab:sysparams}
\end{deluxetable}

Detecting optical auroral emission from the possible atmosphere of Proxima Cen b is likely much more favorable for this system than for an Earth-Sun analog. This is due to both planetary and stellar characteristics that favor auroral production and improve detectability (see Table~\ref{tab:sysparams}). In particular, Proxima Cen b's intrinsic planetary properties may favor production of aurorae from oxygen atoms. If Proxima Cen b is Earth-like in composition, recent dynamical/planetary interior modeling results by \citet{Barnes2016} and \citet{Zuluaga2016} suggest that the planet may have a magnetic field, potentially increasing the likelihood of atmospheric retention and of auroral emission. Atmospheres rich in oxygen-bearing molecules, including O$_2$ and CO$_2$, have been predicted for Proxima Cen b \citep{Meadows2016} as a result of the evolutionary processes for terrestrial planets orbiting M dwarfs \citep{LugerBarnes2015,Barnes2016}. On Earth, the oxygen (OI) auroral line at 5577\AA\ provides the distinctive green glow observed in both the Aurora Borealis and the Aurora Australis, and is the brightest (i.e., highest photon emission rate) auroral feature \citep{Chamberlain1961, Dempsey2005}. For emissions from the upper atmosphere, only the 1.27$\mathrm{\mu}$m O$_2$ airglow and combined near-infrared OH night glow features are brighter \citep{Hunten1967}. The oxygen green line is seen in both the Earth's O$_2$-rich atmosphere \citep{Chamberlain1961} and Venus' CO$_2$-dominated atmosphere \citep{Slanger2001}, where it has been observed to increase in brightness after CME events \citep{Gray2014}.

The stellar properties and the planet-star separation are also likely to enhance the auroral power on Proxima Cen b relative to an Earth-Sun analog. Proxima Centauri is an active flare star with a magnetic field ${\sim} 600\times$ stronger than that of the Sun \citep{Reiners2008,Davenport2016}. Since stellar activity drives auroral emission for an Earth-like magnetosphere, such features may be much stronger on planets orbiting active M dwarfs. Additionally, with a close-in orbit of 0.0485 AU, Proxima Cen b is about $20\times$ closer to Proxima Centauri than the Earth is to the Sun \citep{Anglada-Escude2016}. This proximity further increases particle fluxes incident on the planetary atmosphere that drive ionization and the subsequent recombination radiation.

In addition to increasing the likelihood and strength of the aurora, the characteristics of the Proxima Centauri system may also enhance its detectability. Since the Proxima system is only 1.3pc away, it is perhaps the best-case scenario for the detection of the faint auroral signal from a terrestrial exoplanet. Even though the planet-star contrast ratio in reflected visible light is poor \citep[${\lesssim} 10^{-7}$; see][]{Turbet2016,Kreidberg2016,Meadows2016}, if Proxima Cen b exhibits auroral emission, this will brighten the planet and potentially boost the planet-star contrast by one or more orders of magnitude at the wavelengths of the auroral emission features. The short wavelength of the oxygen green line also improves the contrast of the planet relative to the star due to the star's cool temperature and TiO absorption, which strongly suppresses the brightness of the star in the visible. This improvement in contrast is significantly less for the near-infrared O$_2$ 1.27$\mathrm{\mu}$m and OH airglow lines. In addition to increasing the contrast, the small semi-major axis of Proxima Cen b results in an orbital velocity of ${\sim} 50$ km/s, which will cause its auroral emission to be Doppler-shifted by as much as 1\AA\ over the course of its orbit, making it easier to disentangle it from stellar features via high resolution spectroscopy. An additional advantage of the short wavelength of the OI feature is the smaller inner working angle and point-spread function that may be achieved with a coronagraph at that wavelength \citep{Agol2007}. These factors all improve the chance of detection with ground-based telescopes.

The detection of the oxygen auroral line at 5577\AA\ would provide an important diagnostic for planetary properties. Its detection would not only confirm the existence of the planet, but would point to the presence of an atmosphere with abundant oxygen atoms, which is more likely to indicate a terrestrial body. Additionally, the detection of the line would yield a measurement of the radial velocity (RV) of the planet, which combined with the RV measurements of the star \citep{Anglada-Escude2016} would enable the measurement of the eccentricity and inclination of the orbit, ultimately yielding the mass of the planet \citep[see, e.g.,][]{LovisFischer2010}. Detection of the oxygen auroral line would therefore provide several key planetary parameters that could be used to constrain Proxima Cen b's potential habitability \citep{Barnes2016,Meadows2016}.

This paper is organized as follows: in \S\ref{sec:signal} we calculate the expected auroral emission strength of Proxima Cen b under different assumptions of stellar and planetary properties. In \S\ref{sec:detect} we model the planet-star contrast ratio in a narrow band centered on the OI 5577\AA\ line and calculate the integration times required to detect the feature with different instruments. In \S\ref{sec:search} we conduct a preliminary search for auroral emission in the HARPS high-resolution, ground-based spectroscopy used by \citet{Anglada-Escude2016} for the RV detection of Proxima Cen b. Finally, in \S\ref{sec:disc} we discuss our results and present our conclusions.

\section{Auroral Signal Strength}
\label{sec:signal}

Below, we quantitatively estimate the auroral intensity for steady-state stellar input. We assume the planet to be terrestrial with the orbital characteristics of Proxima Cen b (see Table~\ref{tab:sysparams}) and calculate the auroral emission via two different methods. Method 1 (\S\ref{sec:signal_m1}) involves a simple estimation of the emitted electromagnetic auroral power driven by the stellar wind power delivered at the magnetopause of the planet. Method 2 (\S\ref{sec:signal_m2}) uses the prediction of a magnetohydrodynamical (MHD) model that was tuned to calculate the auroral response at Earth, with modifications to the relevant inputs of the stellar wind of Proxima Centauri and assumed planetary parameters for Proxima Cen b \citep{Anglada-Escude2016}. 

The quantities we calculate include only the estimated, localized emissions caused by magnetospheric particle precipitation into a discrete auroral oval --- not the diffuse, global phenomenon of airglow. On Earth, the 5577\AA\ airglow can be visible to the naked eye and could be significant on Proxima Cen b, but is commonly driven by different physical processes (e.g., nightside recombination due to dayside photoionization) that are outside the scope of this analysis. Similarly, the 5577\AA\ airglow has been observed at Venus \citep[e.g.][]{Slanger2001} and Mars \citep[e.g.][]{Seth2002} --- both having no present-day global magnetic field. For these reasons we cannot suggest basing the existence of or placing constraints on Proxima Cen b's planetary magnetic field based on the detection of this auroral line \citep[see, for instance,][]{Griessmeier2015}. A search for radio emission from Proxima Cen b --- which may be correlated with optical auroral emission, as it is on Earth and on Saturn \citep{Kurth2005} --- would likely be necessary to constrain the planetary magnetic field. However, it is worth noting that an Earth-like 1 kR (1 R = 1 Rayleigh $\equiv 10^{6}\ \mathrm{photons\ s^{-1}\ cm^{-2}}$) airglow across the entire planet would still emit $\sim$2 orders of magnitude less energy than the discrete polar aurora --- see \S\ref{sec:signal_summary} below.

\subsection{Stellar winds at Proxima Cen b}
\label{sec:stellar_winds}

M dwarf mass-loss rates, and therefore stellar winds, are not well constrained due to observational sparsity and difficulty \citep[e.g.][]{Wood2004}. To model the M dwarf winds for Proxima Centauri, we adopt the predictions from the modeling efforts of \citet{Cohen2014}, who generated an MHD stellar wind model for the M3.5 star EV Lacertae based on available observations. There are two primary differences between EV Lac and Proxima Centauri that we should take into account when considering the stellar wind at our planet's location of interest: 1) the relative mass-loss rates, 2) the difference in rotation rates. 

The first of these factors has been estimated by \citet{Wood2005}, who find that the mass-loss per unit surface area for Proxima Centauri and EV Lacertae are quite similar. This suggests comparable wind conditions at equal distances in units of their respective stellar radii. 

The second factor, the rotation rate, affects the morphology of the stellar wind magnetic field by changing the Alfv\'{e}n radius.  The Alfv\'{e}n radius, $R_A$, is defined as the point where the Alfv\'{e}n Mach number is equal to unity --- i.e., $M_A \equiv$ $u_{sw}$/$v_{A}$=1, where $u_{sw}$ is the stellar wind speed and $v_A$ is the Alfv\'{e}n speed. Interior to $R_A$ (the sub-Alfv\'{e}nic wind) the magnetic field of the star is mostly radial, and corotates at the angular rate of the star; exterior to $R_A$ (the super-Alfv\'{e}nic wind) the field begins to lag behind corotation as the magnetic tension is overcome by the flow of the wind. In the super-Alfv\'{e}nic regime, the interplanetary magnetic field (IMF) exhibits the well-known Parker-spiral \citep{Parker1958}. The Alfv\'{e}n point is an important boundary that modifies the energy transfer between the stellar wind and the planetary magnetosphere.

To correctly estimate the interactions, it is important to consider Proxima Cen b's orbital distance from its host star, for both the dynamic parameters (mass density, velocity) and the magnetic structure --- i.e., we must consider where Proxima Cen b orbits relative to its Alfv\'{e}n radius, $R_A$. We note that the rotational period of Proxima Centauri \citep[82.6 days;][]{Collins2016} is $\sim$19 times lower than EV Lacertae \citep[4.376 days;][]{Testa2004}. For our purposes, we estimate an average $R_A$ for a simple stellar dipole moment:
\begin{align}
    R_A = \left( \frac{4 \pi \mathcal{M_\star}^2}{\dot{M}_\star \omega_\star \mu_0} \right)^{\frac{1}{5}},
\end{align}
\noindent where $\mathcal{M_\star}$ is the magnetic dipole moment for the star, $\dot{M}_\star$ is the mass-loss rate, $\omega_\star$ is the angular frequency of stellar rotation, and $\mu_0$ is the vacuum permeability. For EV Lacertae and Proxima Centauri, $R_A$ are $\sim$65.4~$R_\star$ (0.075~AU) and 115~$R_\star$ (0.192~AU), respectively. This is the average value for a simple dipole moment, as we are not including magnetic topology, but nonetheless the value obtained for EV Lac agrees well with the approximate average for the more complicated magnetic treatment simulated in \citet{Cohen2014}. The relative orbit for Proxima Cen b is therefore $\sim$0.76 $R_A$. Coincidentally, this corresponds well to the simulated Planet B at EV Lac in \citet{Cohen2014}, which orbits at $\sim$0.79 $R_A$. 

Recently, \citet{Garraffo2016} applied an MHD model of stellar winds based on the Zeeman-Doppler Imaging (ZDI) of GJ51, and scaled the magnitude of the surface field to match the anticipated value of 600 G for Proxima Centauri. Their results from the assumed magnetic environment are in line with the values we adopt from Table~\ref{tab:swparams}, and our value calculated for the magnetopause distance using Eq.~\ref{eq:pressbal} below is within the range of their calculations for magnetopause distance for Proxima Cen b. However, the structure of the magnetic topology in the simulation of \citet{Garraffo2016} places Proxima Cen b primarily in the super-Alfv\'{e}nic wind, contrary to both the simple method above and the bulk of the structure found by \citet{Cohen2014}.

Our estimate of $R_A$ does not take into account the complicated magnetic topology of a realistic stellar magnetic field, which could indicate the planet likely orbits primarily through sub-Alfv\'{e}nic conditions \citep[e.g., Fig. 1 of][]{Cohen2014} or through primarily super-Alfv\'{e}nic conditions \citep[Fig. 2 from][]{Garraffo2016}. Therefore, we consider both super- and sub-Alfv\'{e}nic conditions for the steady-state stellar wind, using the reported parameters at Planet B from \citet{Cohen2014}; see Table~\ref{tab:swparams}.
 
\begin{deluxetable}{ccccc}
\tablewidth{\linewidth}
\tablecaption{Stellar wind conditions}
\tablenum{2}
\tablehead{\colhead{Quantity} & \colhead{Sub-Alfv\'{e}nic} & \colhead{Super-Alfv\'{e}nic} } 
\startdata
$n$ (cm$^{-3}$) & 433 & 12895  \\
$T$ (10$^5$ K) & 3.42 & 4.77 \\
{\bf u} (km s$^{-1}$) & (-630, -1, 30) & (-202, 102, 22) \\
{\bf B} (nT) & (-804, 173, 63) & (-57, 223, 92) \\
$M_{A}$ & 0.73 & 4.76
\enddata
\tablecomments{Stellar wind conditions from \citet{Cohen2014}, at EV Lacertae for $a{\sim}51.98$ R$_{*}$ (0.073 AU). $n$ is the stellar wind number density, $T$ is the ion temperature, {\bf u} is the velocity, {\bf B} is the interplanetary magnetic field (IMF), and $M_{A}$ is the Alfv\'{e}n mach number. \label{tab:swparams}}
\end{deluxetable}

\subsection{Magnetic dipole moment of Proxima Cen b}
\label{sec:dipole_moment}
Tidal locking is likely for the expected orbital parameters of Proxima Cen b and the age of the system. We therefore expect a rotational period equal to the orbital period, 11.186 days, or 8.94\% of the Earth's rotational frequency. Following the magnetic moment scaling of \citet{Stevenson1983} and \citet{Mizutani1992}, we assume the upper limit of the rotationally-driven planetary dynamo as $\mathcal{M} \propto \omega^{1/2} r_c^3$, where $\mathcal{M}$ is the magnetic moment, $\omega$ is the rotation rate of the planet, and $r_c$ is the core radius (which we assume to be proportional to the planetary radius). This suggests a magnetic moment for an Earth-radius Proxima Cen b of ${\sim}0.3 \mathcal{M}_{\oplus}$. Taking the upper limit of the expected radius of Proxima Cen b, 1.4 R$_{\Earth}$, this gives a magnetic moment of ${\sim}0.8 \mathcal{M}_{\oplus}$, which agrees with the upper limit of \citet{Zuluaga2016}. However, \citet{Driscoll2015} showed that for an Earth-like terrestrial planet orbiting a star of 0.1 M$_{\sun}$ with high initial eccentricity ($e\geq$0.1) within 0.07 AU, the planet will circularize before 10 Gyr. On this timescale, the orbital energy dissipated as tidal heating is sufficient to drive a strong convective flow in the planetary interior that could generate a magnetic moment in the range of ${\sim}0.8-2.0 \mathcal{M}_{\Earth}$ during the process of circularization. Given the above, we consider the situation of an Earth magnitude magnetic field for Proxima Cen b, but discuss how each of the methods below can be scaled to various magnetic dipole moments.

\subsection{Auroral stellar wind power scaling}
\label{sec:signal_m1}

%% Fig: Method 1 - stellar wind scaling %%
\begin{figure}[bt]
\includegraphics[width=0.47\textwidth, angle=0]{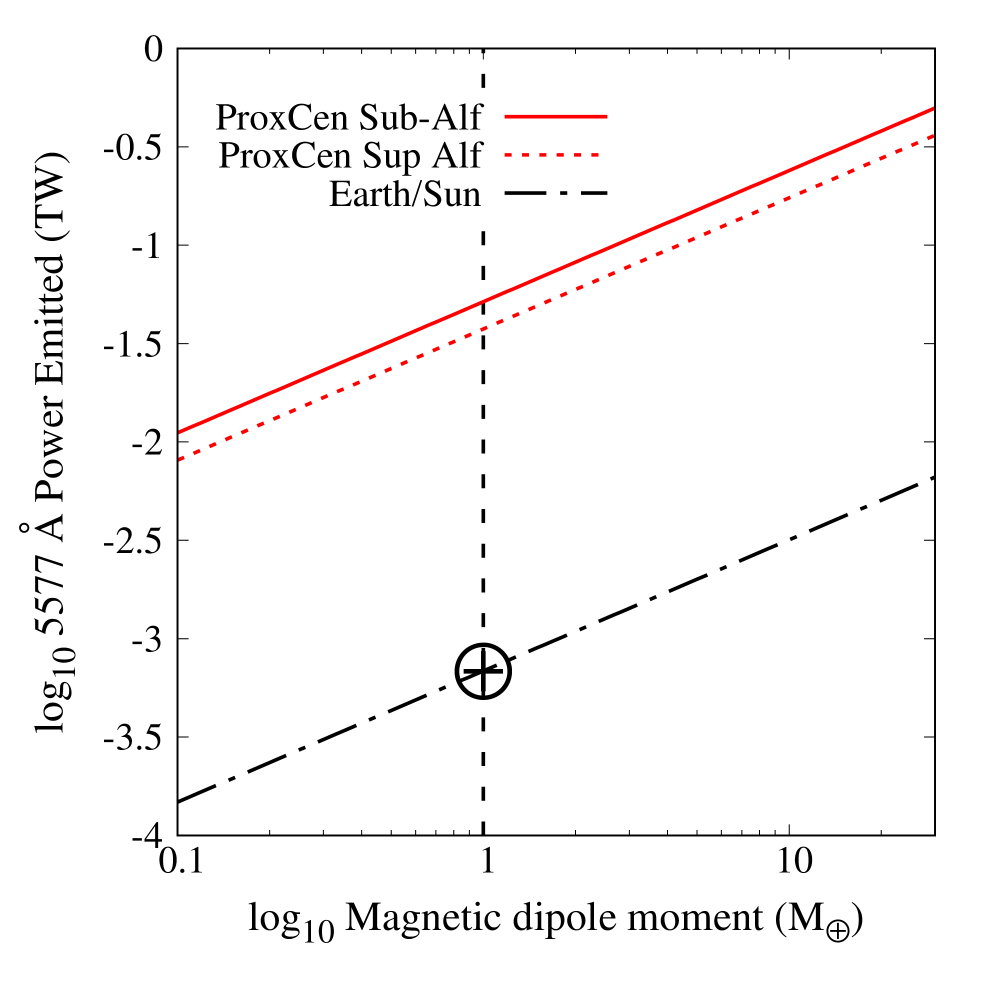}
\caption{Predicted 5577\AA\ auroral power as a function of planetary magnetic dipole moment calculated using the stellar wind scaling method from \S\ref{sec:signal_m1}. The solid (dotted) red line corresponds to the sub-\mbox{(super-)} Alfv\'{e}nic stellar wind conditions at Proxima Cen b. The black dash-dotted line corresponds to Earth in its natural orbit around the Sun, and the black Earth symbol corresponds to the method's calculation for Earth. The dashed vertical black line indicates an Earth-equivalent magnetic dipole moment.\\[0in]}
\label{fig:auroral_power}
\end{figure}

\citet{Desch1984} suggested a correlation between incident stellar wind power and the power of planetary radio emissions in the solar system, a so-called ``radiometric Bode's law.'' \citet{Zarka2006,Zarka2007} extended the work to modern solar system measurements as well as potential exoplanetary systems, and further suggested that a similar ``auroral UV-magnetic Bode's law'' could exist, though the author notes such scaling would be less generally applicable than the radio case across planetary systems due to the complexities of UV auroral generation for differing planetary atmospheres and magnetospheric dynamics. The calculations in this section can be thought of similarly as a ``visible-kinetic Bode's law'' for the specific case of exoplanets with an Earth-like atmosphere. A similar relation may also be derived for the magnetic stellar wind interaction (e.g., a ``visible-magnetic Bode's law''; details below).

The stellar wind kinetic power delivered to the magnetosphere of the planet can be expressed as:
\begin{align}
    P_{U} = \rho\,v^3\,\pi\,R_{MP}^2, \label{eq:swpower}
\end{align}
where $\rho$ and $v$ are the stellar wind mass density and velocity relative to planetary motion ($\sim$48 km/s), respectively, and $R_{MP}$ is the magnetopause distance along the line connecting the star and planet (sub-stellar point). The latter can be estimated through magnetospheric pressure balance with the stellar wind dynamic ram pressure \citep[e.g.][]{Schield1969}:
\begin{align}
    \frac{2\,\mathcal{M}^2}{K_{SW} \mu_0\,R_{MP}^6} = p_{ram}, \label{eq:pressbal}
\end{align}
where $\mathcal{M}$ represents the magnitude of the magnetic dipole moment, K$_{SW}$ is related to particle reflection at the magnetopause (herein the interaction is assumed to consist of inelastic collisions, or K$_{SW}$=1), and $R_{MP}$ is the distance from the planet at which the magnetic pressure of the planet balances the pressure of the stellar wind. The RHS of Eq.~\ref{eq:pressbal} represents the dynamic ram ($p_{ram}=\rho v^2$) pressure of the stellar wind, calculated from the values in Table~\ref{tab:swparams}.

We consider an Earth-strength magnetic dipole moment of $\mathcal{M}=8.0\times$10$^{15}$ Tesla m$^3$. Solving for $R_{MP}$ in Eq.~\ref{eq:pressbal} and inserting into Eq.~\ref{eq:swpower} provides an estimate of the stellar wind power incident on the planetary magnetopause. Externally-driven planetary auroral systems are not typically 100\% efficient at converting the incident stellar wind power into electromagnetic auroral emission, and range from $\sim$0.3\% at Neptune, $\sim$1\% at Earth, and up to almost 100\% at Jupiter \citep[e.g.][]{Cheng1990,Bhardwaj2000}. For reference, at Earth, this method gives us a reasonable estimate of the total emitted electromagnetic auroral power of $\sim$30 GW for nominal solar wind conditions (4 cm$^{-3}$, 400 km s$^{-1}$), which is consistent with the anticipated power of 1-100 GW, depending on solar and magnetospheric activity. While the intensities of various emissions vary widely with activity and atmospheric conditions, we assume an averaged auroral emission. In order to estimate the emitted power of the OI 5577\AA\ line, we assume it represents 2\% of all emitted electromagnetic power \citep{Chamberlain1961,Kivelson1995}, as calculated by Eq.~\ref{eq:swpower} We note that this assumes an Earth-like atmosphere for the planet; we briefly discuss the effect of different atmospheric compositions in \S\ref{sec:disc}.

Fig.~\ref{fig:auroral_power} shows the predicted emitted power of the 5577\AA\ line based on Eq.~\ref{eq:swpower} and multiplied by the 2\% factor mentioned above and by the conversion efficiency of 1\%. For the Earth, this method predicts a power of $\phi_{\Earth}\sim$0.68 GW in the 5577\AA\ line. Assuming a 5$^{\circ}$ latitudinal width starting at $\sim$18$^{\circ}$ co-latitude and extending equatorward, this corresponds to a photon flux of $\sim$13.7~kR. This is in agreement with moderate auroral activity \citep[IBC II\footnote{IBC = International Brightness Coefficients, a standardized scale for quantifying auroral intensities \citep[see, e.g.,][]{Hunten1955}.}, 10 kR 5577\AA\ emission; see Table II.1 in][]{Chamberlain1961}, and within a factor of 2--5 of observations during moderate geomagnetic disturbance \citep[2.5--6 kR 5577 \AA\ emission, e.g.][]{Steele1990}.

Power estimates for the 5577\AA\ line for a 0.05 AU orbit around Proxima Centauri are shown in Table~\ref{tab:auroral_power}. The calculated power is $\sim$75.3 (54.7) times $\phi_{\Earth}$, in the sub-\mbox{(super-)} Alfv\'{e}nic stellar wind. These are the estimates for a steady-state stellar wind, for a terrestrial planet with an Earth-like magnetic dipole moment. Note that by inspection of Eqs.~\ref{eq:swpower} and~\ref{eq:pressbal}, one can see that the expected power scales as $\mathcal{M}^{2/3}$, and so can easily be extended to different planetary dipole moments.

The method above has a weakness in that it completely ignores the incident Poynting flux from the IMF, and potential direct magnetic interactions between the stellar wind and planetary magnetic field, e.g. flux merging or reconnection. These interactions can produce a significant amount of magnetospheric energy input, and so they are important to consider. Similar to Eq.~\ref{eq:swpower}, a scaling relation between power emitted at the 5577 \AA\ line and incident magnetic flux in the stellar wind, akin to a visible-magnetic Bode's law, can be given as \citep[e.g.,][]{Zarka2006,Zarka2007,Grie2007}:
\begin{align}
    P_{B} = \epsilon\, K\, \left( \frac{v\,B_{\perp}^2}{\mu_0} \right)\,\pi\,R_{MP}^2 \label{eq:swpowerB}
\end{align}
where $\epsilon$ is the efficiency of reconnection (typically of order 0.1--0.2), $K$ is related to the ``openness'' of the magnetosphere, and for an Earth-like dipole is $K = \sin^4(\theta/2)$ where $\theta$ is the angle between the perpendicular IMF and planetary dipole field, $B_{\perp}$ is the perpendicular IMF ($\sqrt{B_Y^2+B_Z^2}$), $\mu_0$ is the vacuum permeability, and $R_{MP}$ is the magnetopause sub-stellar point discussed above. We can estimate the magnetic interaction at Proxima Cen b using our stellar wind conditions by taking the ratio of Eqs.~\ref{eq:swpowerB} and~\ref{eq:swpower}:
\begin{align}
    \frac{P_B}{P_U} = \frac{\epsilon\, K\, B_{\perp}^2}{\mu_0\, \rho\,v^2}, \label{eq:swpower_ratio}
\end{align}
which is essentially the ratio of the perpendicular IMF magnetic pressure to the ram pressure, modulated by magnetic field orientation and reconnection efficiency. In the best case scenario, $K$ is equal to $1$ (indicating $\theta$=$\pi$, driving strong reconnection at the magnetopause), and $\epsilon$ is of order 0.2 or so. Assuming this best case, and inserting the values from Table~\ref{tab:swparams} for the sub- and super-Alfv\'{e}nic cases, one obtains a ratio of $\sim$0.019 and 0.0084 for the sub- and super-Alfv\'{e}nic cases, respectively. For the particular stellar wind parameters we have chosen, the kinetic power dominates the anticipated auroral output for Proxima Cen b. It is worth noting, however, that the magnetic environment (both planet and star) is largely unconstrained, and highly dynamic---particularly near active M dwarfs.

\begin{deluxetable*}{lccccc}

\tablewidth{\linewidth}
\tablecaption{Calculated 5577\AA\ auroral power, by method}
\tablenum{3}
\tablehead{\colhead{Case} & \colhead{Method 1} & \colhead{Method 2 (quiet)} & \colhead{Method 2 (SS)} & \colhead{Method 2 (CME)} & \colhead{Method 2 (CME+SS)} \\
\colhead{} & \colhead{[TW]} & \colhead{[TW]} & \colhead{[TW]} & \colhead{[TW]} & \colhead{[TW]} } 
\startdata
Prox Cen b (Sub) & 0.051 & 0.09 & 0.24 & 8.103 & 21.42 \\
Prox Cen b (Sup)& 0.038 & 0.049 & 0.14 & 4.41 & 12.10\\
Earth/Sun & 6.7$\times$10$^{-4}$ & 7.5$\times$10$^{-4}$ & 1.5$\times$10$^{-3}$ & 0.068 & 0.1317
\enddata
\tablecomments{Power emitted for the OI 5577\AA\ line in terrawatts (TW) for an Earth-strength magnetic dipole on Proxima Cen b in the sub-Alfv\'{e}nic (Sub) and super-Alfv\'{e}nic (Sup) stellar winds. For method 2: column 2 assumes no significant stellar activity and a quiet magnetosphere; column 3 assumes geomagnetic substorm (SS) activity; column 4 assumes CME conditions in the stellar winds, but no magnetospheric disturbance; column 5 assumes both CME conditions and substorm activity. \label{tab:auroral_power}}

\end{deluxetable*}

\subsection{3D MHD empirical energy coupling}
\label{sec:signal_m2}
 
Eqs.~\ref{eq:swpower}~\&~\ref{eq:swpowerB} above are decent first approximations, but involve significant uncertainties concerning the energy dissipation in physical phenomena throughout the magnetosphere (i.e., auroral activity) \citep{Perreault1978,Akasofu1981}. \citet{Wang2014} developed a global, 3D MHD model to obtain a fit for the energy coupling between the solar wind and Earth's magnetosphere to estimate the energy transferred directly from the wind into the magnetosphere and auroral precipitation (see their Eq.~13, and below). The simulations were performed over 240 iterations across their solar wind parameter space, and the resulting nonlinear fit for the energy transfer to the terrestrial magnetosphere was found to be
\begin{align}
    P_{trans} = K_1\,n_{sw}^{0.24}\,v_{sw}^{1.47}\,B_T^{0.86}\,\left[\sin^{2.7}(\theta/2)+0.25 \right] \label{eq:mhd_fit},
\end{align}
where $K_1=3.78\times 10^7$ is a coupling constant, $n_{sw}$ and $v_{sw}$ are the stellar wind number density (in cm$^{-3}$) and velocity relative to planetary motion (in km s$^{-1}$), respectively, $B_T$ is the magnitude of the transverse component of the Sun's IMF ($B_T = \sqrt{B_X^2+B_Y^2}$) in nT, and $\theta$ is the so-called IMF clock angle ($\tan\theta = B_Y/B_Z $). The coordinate system used is the geocentric solar magnetospheric (GSM) system, with $\hat{X}$ pointing from the planet to the star, $\hat{Z}$ aligned with the magnetic dipole axis of the planet (here assumed to be perpendicular to the ecliptic), and $\hat{Y}$ completing a right-handed coordinate system.

\citet{Wang2014} were focused solely on the Earth's magnetosphere, but one can scale to any dipole moment by noting that Eq.~\ref{eq:mhd_fit} scales just as in \S\ref{sec:signal_m1}: the dipole moment term is implicitly included in the coupling constant $K_1$ above and scales with the planetary magnetic dipole magnitude as $\mathcal{M}_P^{2/3}$ \citep[][also Eqs.~\ref{eq:swpower} \& \ref{eq:pressbal} above]{Vasyliunas1982}.
 
Eq.~\ref{eq:mhd_fit} is the total power delivered by the stellar wind to the magnetosphere, which \citet{Wang2014} estimate is $\sim$13\% of the total incident stellar wind energy. They further estimate that 12\% of that energy is dissipated by particle precipitation into the auroral regions, yielding a total solar wind/auroral coupling efficiency of $\sim$1.56\% -- very similar to the efficiency value of 1\% assumed for Earth and Proxima Cen b in \S\ref{sec:signal_m1}. As simple validation for our purposes, we use this method to predict a maximum coupling of auroral particle precipitation (with IMF clock angle $\theta =$ $\pi$, driving reconnection and likely substorm activity) at Earth of $\sim$0.17 TW. This is in agreement with terrestrial plasma observations during periods of geomagnetic disturbance \citep[e.g.][]{Hubert2002}. This method is useful in that it provides a direct relationship between the power delivered as auroral particle precipitation and incident stellar wind conditions.

For Proxima Cen b subjected to the stellar winds from Table~\ref{tab:swparams}, this method predicts a total power of auroral particle precipitation of ${\sim} 10.7$ (5.8) TW for the sub-\mbox{(super-)}Alfv\'{e}nic stellar wind. The stellar wind parameters in Table~\ref{tab:swparams}, however, are a snapshot and not indicative of the highly variable conditions likely experienced at Proxima Cen b.

Magnetospheric substorms, related to transient populations of energized particles driven by magnetic reconnection in the magnetotail, can drive strong increases in auroral particle precipitation. Though not a one-to-one indicator, substorm activity can be associated with periods of strong reconnection at the magnetopause---correlated with a significant negative $B_Z$ component in the IMF. In the present work, we assume $\theta$=$\pi$, or B$_Y$=0, to obtain an upper limit to substorm influence under our model. Although this is not a strict definition, \citet{Wang2014} calibrated the model used here to include periods of substorm activity and high hemispheric energy input. Assuming with this strong negative $B_Z$ that a substorm is driven at Proxima Cen b, we predict an energy input of $\sim$28.3 (15.9) TW for the sub-\mbox{(super-)}Alfv\'{e}nic wind. To compare directly to the 5577\AA\ line auroral power output such as that calculated in \S\ref{sec:signal_m1}, we must link these values to the aurora by including the efficiency of precipitating charged particles in the production of auroral emission for the 5577\AA\ line, which will be done below.

To calculate the auroral 5577\AA\ photon flux, we use the precipitating auroral particle powers above obtained from Eq.~\ref{eq:mhd_fit}, and combine with the anticipated size of the auroral oval and an observed conversion efficiency for electron precipitation to 5577\AA\ emission. This gives the photon flux in kR, $\phi_{5577}$:
\begin{align}
    \phi_{5577} = P_{in}\, A_{mag}^{-1}\, \epsilon_e,
\end{align}
where $P_{in}$ is 12\% (discussed above) of P$_{trans}$ from Eq.~\ref{eq:mhd_fit}, $A_{mag}$ is the summed area of both the northern and southern auroral ovals (we assume N-S symmetry), and $\epsilon_e$ is the efficiency with which magnetospheric electrons are converted to auroral emission of the 5577\AA\ oxygen line. We use the reported values from \citet{Steele1990} (noted below), who used ground-based observations of auroral line intensities and the related satellite observations of energetic electron flux to draw a relation between electron precipitation and auroral photon flux. We then integrate the resulting flux over a nominal 5$^\circ$ auroral oval (for each hemisphere), the colatitude of which is dependent on the sub-stellar magnetopause distance (discussed below). 

\citet{Steele1990} reported the conversion efficiency for the 5577\AA\ OI line as 1.73$\pm$0.51 (1.23$\pm$0.44) kR/(erg cm$^{-2}$ s$^{-1}$) for a magnetospheric Maxwellian electron population of characteristic temperature 1.8 (3.1) keV. In the present work, we take the average values for these populations, ${\sim} 1.48$ kR/(erg cm$^{-2}$ s$^{-1}$). We assume the fraction of total hemispheric power ($P_{in}$) delivered by electrons to be 0.8 \citep{Hubert2002}, so this factor is included in the P$_{in}$ factor.

The magnetopause distance we calculate via Eq.~\ref{eq:pressbal} for the Earth-like magnetic dipole moment is $\sim$4.2 (3.3) $R_P$ for the (sub-)super-Alfv\'{e}nic conditions. From these values, we can provide a simple estimate of the total auroral oval coverage. The magnetic co-latitude of the boundary between open and closed flux for our assumed ideal planetary dipole geometry (i.e., the co-latitude where the field structure no longer intersects the planetary surface) is sin$^{-1}$(1/$\sqrt{R_{MP}}$) \citep{Kivelson1995}. Discrete auroral activity occurs primarily due to energized plasma originating from closed field structure stretched out behind the planet in the stellar wind, i.e., the magnetotail. This field structure intersects the planet equatorward of the open/closed boundary co-latitude. If we assume a nominal 5$^\circ$ auroral oval width beginning at the co-latitude obtained, and extending equatorward, we calculate a single-hemisphere coverage of $\sim$1.17$\times$10$^{17}$ cm$^2$ for the auroral oval under sub-Alfv\'{e}nic conditions, and $\sim$1.30$\times$10$^{17}$ cm$^2$ under super-Alfv\'{e}nic conditions.

Following the above, we obtain a photon flux value of $\phi_{5577}$ $\sim$2.26 (1.16) MR for the sub-\mbox{(super-)}Alfv\'{e}nic wind conditions. This corresponds to our predicted emission power in Table~\ref{tab:auroral_power}, method 2 (quiet) of $\sim$0.090 (0.049) TW  under steady-state sub-\mbox{(super-)}Alfv\'{e}nic conditions. For the maximum emission during a magnetospheric substorm, we obtain values of $\sim$0.24 (0.14) TW for sub-\mbox{(super-)}Alfv\'{e}nic winds.
 
There is another case of interactions that we should consider that involves stellar activity --- flaring and coronal mass ejections (CME). During these events, stellar wind densities could increase by a factor of $\sim$10, velocities by a factor of $\sim$3, and IMF magnitude by a factor of $\sim$ $10-20$ \citep{Khodachenko2007,Gopalswamy2009}. Inserting such ratios in the 3D MHD-fit predicted power in from Eq.~\ref{eq:mhd_fit}, we predict transient maximum 5577\AA\ emissions of $\sim$8.10 (4.40) TW for the sub-\mbox{(super-)}Alfv\'{e}nic CME conditions. For the maximum emission during a magnetospheric substorm under CME conditions, we obtain values of $\sim$21.42 (12.10) TW for sub-\mbox{(super-)}Alfv\'{e}nic winds. These transient CME conditions can have timescales of ${\sim} 10-10^3$ minutes per event, with multiple, consecutive events possible. Given that \citet{Davenport2016} report such high stellar activity for Proxima Centauri, Proxima Cen b could experience CME impacts for a large percentage of its orbital phase \citep[e.g.,][]{Khodachenko2007}. 

\subsection{Unmagnetized planet}
\label{sec:unmagnetized}

The above results all assume a large, Earth-like planetary magnetic dipole moment for Proxima Cen b. If, in fact, the planet does not sustain a global dynamo, it will only be protected by a relatively thin spherical shell ($\sim$1000 km) of plasma in the upper atmosphere - similar to Earth's ionosphere.

For the sub-Alfv\'{e}nic case, the interaction is a unipolar interaction similar to the Jupiter-Io interaction \citep{Zarka2007}. In this case, the power dissipated by the wind is similar to the form of Eq.~\ref{eq:swpowerB}, where $\epsilon$ and $K$ are replaced by a single parameter indicating the fraction of magnetic flux convected onto the ``obstacle'' (the ionosphere), and $R_{MP}$ becomes the size of the ``obstacle'' --- for an Earth-like ionosphere, ${\sim}1.16$ planetary radii. While there is no dipolar focusing mechanism for the particle precipitation in this case, it is worth considering the energized particles flowing on the flux tube connecting the unmagnetized planet with the star, producing maxima on the unmagnetized body in the plane perpendicular to the IMF \citep[see, e.g.,][]{Saur2000,Saur2004}.

Assuming 100\% of incident magnetic energy flux is convected onto the planet and ionosphere, the expected 5577\AA\ auroral power becomes 5.9$\times$10$^{-4}$ TW for the sub-Alfv\'{e}nic stellar wind conditions in Table~\ref{tab:swparams}. This interaction is likely insignificant in the context of remote sensing.

For the super-Alfv\'{e}nic flow, this could be considered as analogous to Venus' situation, which sustains no global magnetic field. In this case, the discrete aurora would obviously not be expected due to a lack of magnetic structure, though induced airglow is still a consideration. Lacking a planetary magnetosphere, the magnetic structure fails to focus precipitating particles into the upper atmosphere of such a planet, though there is still magnetic interaction at the planet. The ionosphere is a spherical, conducting shell, and so interacts with the magnetic flux from the IMF as it drapes over and around the planet. Energized particles in the impacting stellar wind magnetic flux could still dip down into the upper atmosphere, depositing sufficient energy to produce airglow - this is especially true for the strong flows from CME activity, or fast stellar wind flow.

A study of the intensity of 5577\AA\ oxygen emission at Venus, relative to Earth, for CME/flare events from the Sun was performed by \citet{Gray2014}. The results indicated that the airglow was relatively on par with that of Earth's upper atmosphere, varying between 10 and a few hundred Rayleigh, which, if integrated over an entire hemisphere of an Earth-like planet, gives a value less than 1\% of the discrete values given in Table~\ref{tab:auroral_power}. For the super-Alfv\'{e}nic flow in Table~\ref{tab:swparams}, the number density is $\sim$1500 times greater than the average at Venus, and the velocity is a factor of $\sim$0.5 that at Venus or Earth. Given the power scales as $\rho\,v^3$, this is a factor of $\sim$200 greater power delivered to the planet. Assuming that airglow at an unmagnetized Proxima Cen b scales linearly with the incident power, this would give a brightness of $\sim$2-60 kR, which is at most a factor of $\sim$5 times the value for Earth using Method 1 and 2 in Table~\ref{tab:auroral_power}, or on the order of 10$^{-3}$ TW.  Given our discussion of auroral detectability below (\S\ref{sec:detect}), we do not expect that this signal could be observed with either current or upcoming missions.

\subsection{Signal Summary}
\label{sec:signal_summary}

The preceding estimates are mostly conservative.  It is possible that all the auroral numbers reported for the sub-Alfv\'{e}nic cases above could be a factor of 4--5 (or more) larger. We are assuming a simple dipolar interaction with the stellar wind, which isn't specifically the case for a planetary dipole in the sub-Alfv\'{e}nic stellar wind; these interactions are more akin to the interactions of Ganymede and Io with the corotating magnetosphere of Jupiter, with the formation of Alfv\'{e}n wings. Modeling efforts by \citet{Preusse2007} showed that for a giant planet with a dipole magnetic moment, field-aligned currents (which are associated with auroral activity) are significantly stronger for planets orbiting inside the Alfv\'{e}n radius of their stellar host. Our estimates, therefore, could be viewed as lower limits. It is also worth noting that \citet{Cohen2014} suggested that a transition between the sub- and super-Alfv\'{e}nic conditions would likely produce enhanced magnetospheric activity and therefore could lead to a periodicity in the auroral activity depending on combined planetary orbital and stellar rotational phases. 
 
For planets in the solar system, only Mars and Earth exhibit observed, significant 5577\AA\ emission for both diffuse airglow and discrete aurora. Mars does not presently have a global magnetic field, but there are crustal regions containing the remnants of previous magnetization that exist and focus particles into the upper atmosphere to produce a relatively weak (inferred $\sim$30 R at 5577\AA) discrete aurora that is $\sim$10 times the strength of the nominal airglow \citep[e.g.][]{Acuna2001,Bertaux2005,Lilensten2015}. On Earth, the airglow and aurora are typically in the range  0.01--1 kR and 1--1000 kR, respectively. During transient periods of minimal auroral activity and maximum airglow emission, emissions can be roughly equivalent, but the average ratio of airglow emission to auroral emission is $\leq$1\%  \citep[e.g.][]{Chamberlain1961,Greer1986}. Even for a constant, planet-wide 1 kR airglow on an Earth-sized planet at Proxima Cen b ($R_P\sim$6371 km), the total signal from the observer-facing hemisphere would be $\sim$4.54$\times$10$^8$W, which is $\sim$1\% of the lowest signal from Table~\ref{tab:auroral_power} and would not be detectable.

Nevertheless, it is important to note that the FUV flux from Proxima Centauri is nearly two orders of magnitude higher than that of the Sun \citep{Meadows2016}.  Airglow stemming from recombination of photodissociated O$_2$ and CO$_2$ could thus be significantly stronger on Proxima Cen b than on Earth. \citet{barthelemy2014} stress the importance of stellar UV/FUV emissions on the production of UV and visible aurorae, and note that, e.g., Lyman-$\alpha$ flux can contribute up to 25\% to the production of the O($^1D$) red-line. However, even if Proxima Cen b had a sustained 100 kR airglow---one hundred times the maximum Earth airglow---its emission would be comparable to the lowest estimate of auroral emission in Table~\ref{tab:auroral_power}, which is still unlikely to be detectable (see \S\ref{sec:detect}). We therefore ignore this potential contribution in the present work, noting that a detailed photochemical treatment would be required to pin down the expected airglow emission at Proxima Cen b.

In summary, we predict a steady-state auroral emission at 5577\AA\ from Proxima Cen b that is of order 100 times stronger than seen on Earth for a quiet magnetosphere, corresponding to an emitted auroral power for the OI line on the order of ${\sim} 0.090\, (0.049)$ TW for the sub-\mbox{(super-)}Alfv\'{e}nic winds using method 2 (\S\ref{sec:signal_m2}). We believe that this method yields more realistic results than the purely kinetic power estimate in method 1 (\S\ref{sec:signal_m1}), due to the inclusion of magnetic interactions in method 2 --- though the magnitudes are similar to within a factor of 2. Assuming Proxima Cen b is an Earth-like terrestrial planet, our maximum transient power estimate for the 5577\AA\ line for CME conditions that drive a magnetospheric substorm is $\sim$21.42~TW, or ${\sim}$30,000 times stronger than on Earth under nominal solar wind conditions. The actual values for Proxima Cen b will naturally change based on planetary parameters (e.g., magnetic dipole moment, magnetospheric particle energy distributions, substorm onset, atmospheric Joule heating) and stellar activity. By our analysis, a ${\sim} 10^3$ (or higher) enhancement compared to Earth as suggested by \citet{OMalley2016} is only possible due to one or more of the following: transient magnetospheric conditions driven by either CME or substorm activity, a magnetic dipole significantly stronger than Earth's, or higher stellar mass-loss than predicted \citep{Wood2005,Cohen2014}.
 
\section{Auroral Detectability}
\label{sec:detect}

In this section we assess the detectability of the 5577\AA\ OI auroral emission line from the atmosphere of Proxima Cen b. Below, we investigate the line profile shape and then calculate planet-star contrast ratios and integration times required for auroral detection.

\subsection{OI Auroral Line Profile}
\label{sec:line_profile}

To estimate the signal-to-noise as a function of spectral resolution, we need to estimate the auroral spectral line width.  The OI $5577$\AA\ green line has no hyperfine structure and because it is a forbidden line, it has negligible ($\lesssim 10^{-15}\mathrm{\AA}$) natural width \citep{Hunten1967}.
Spectroscopic observations of the OI airglow by Keck/HIRES \citep{Slanger2001} and by HARPS \citep[][see \S\ref{sec:search}]{Anglada-Escude2016} are unresolved, revealing the resolution element width of the instrument used for the observation at the wavelength of the line (${\sim}0.1$\AA\ Keck/HIRES; ${\sim}0.05$\AA\ HARPS) rather than the full width at half maximum (FWHM) of the line.

To determine the width of the line, we examine several line broadening mechanisms that play a key role in terrestrial atmospheres.   The planet's rotation will broaden the OI line, but calculations by \citet{Barnes2016} and \citet{Ribas2016} show that Proxima Cen b is likely tidally locked with a rotation period of 11.2 days, resulting in negligible instantaneous rotational broadening ($FWHM = 0.002$\AA).  Pressure broadening can also be safely neglected since OI auroral emission occurs in terrestrial atmospheres at an elevation of ${\sim}100$ km where the atmosphere is thin \citep{Slanger2001}. Similarly, broadening due to atmospheric turbulence can also safely be neglected due to the stratospheric origin of the line. Thermal Doppler broadening should therefore be the dominant line broadening mechanism, resulting in a Gaussian line profile. For the $5577$\AA\ OI line, Doppler broadening gives the following scaling relation:
\begin{align}
    FWHM = 2\Delta \lambda = 0.014 \left ( \frac{\text{T}}{200 \text{ K}} \right )^{1/2} \text{ \AA}, \label{eq:doppler}
\end{align}
where $T$ is the temperature of the emitting layer, for which we adopt the value of 200 K \citep[c.f.][]{Slanger2001}. 
A FWHM of 0.014\AA\ is in good agreement with the Fabry-Perot interferometric line width measurements of \citet{Wark1960}.

\begin{figure}[bt]
\includegraphics[width=0.47\textwidth]{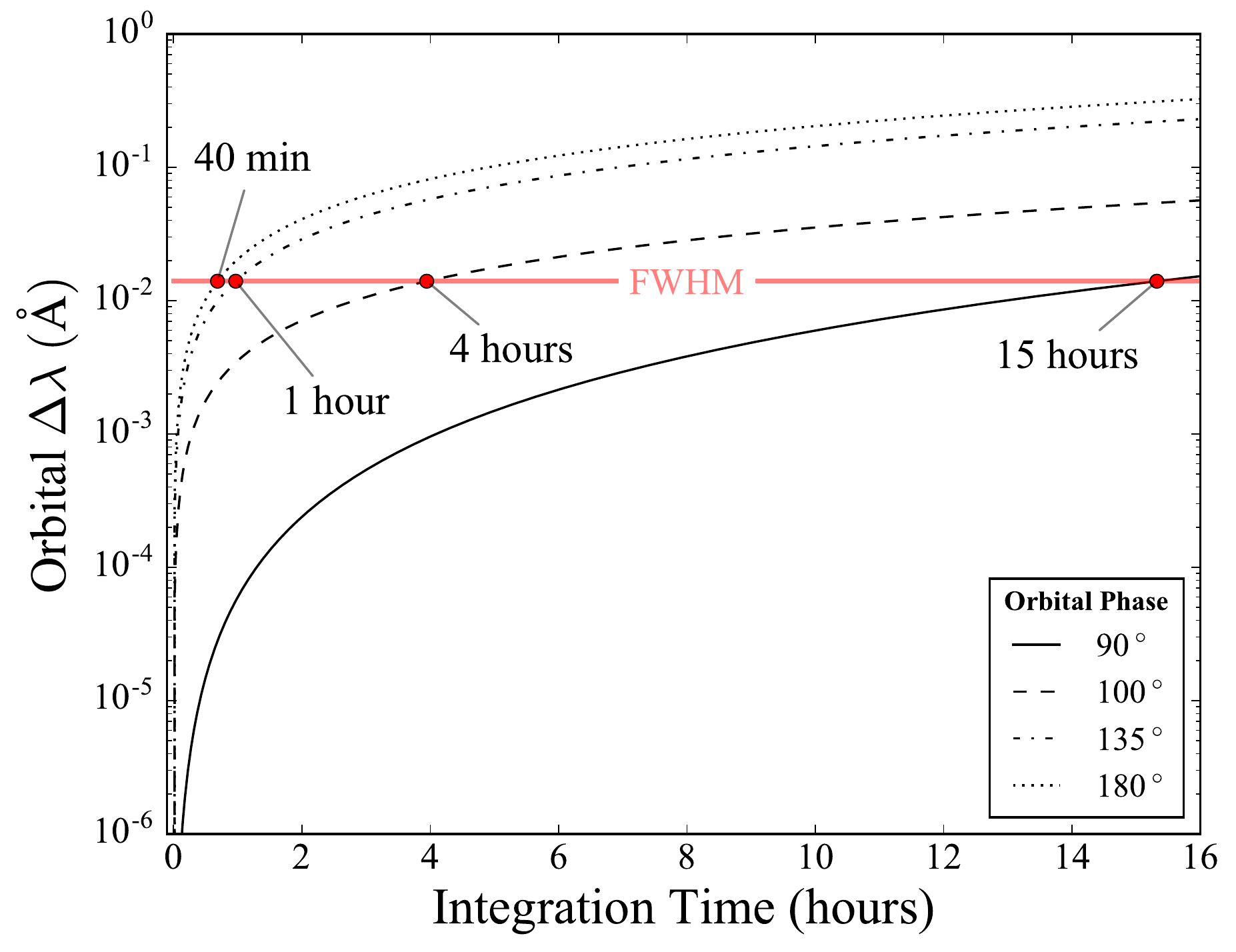}
\caption{Orbital broadening of the 5577\AA\ OI line as a function of the exposure time for observations made at different orbital phases: 90$^\circ$ (quadrature), 100$^\circ$, 135$^\circ$, and 180$^\circ$ (full phase). The FWHM given by Eq.~\ref{eq:doppler} (0.014\AA) is indicated by a horizontal red line; the intersection of this line with the black curves corresponds to the integration time for which the FWHM doubles. At quadrature, exposures up to ${\sim} 6$ hours long have a negligible effect ($\Delta\lambda \lesssim 10^{-3}$\AA) on the width of the line. At all other phases, the broadening is larger and can cause a significant increase in the FWHM in ${\sim} 1$ hour.\\[0in]}
\label{fig:orbital_broadening}
\end{figure}

Given the relatively short period of Proxima Cen b and the long exposure times expected for high resolution spectroscopy, we must also consider the possibility of broadening due to the orbital motion of the planet over the course of an observation. One could take a series of shorter exposures, but this strategy will introduce significant instrumental noise, which is likely to overwhelm any planetary signals. In Fig.~\ref{fig:orbital_broadening} we plot the orbital broadening of the 5577\AA\ line as a function of the exposure time, calculated from the maximum change in the radial velocity of the planet over the course of the observation and assuming an inclination of 90$^\circ$. The effect is strongest at full and new phases (dotted line), where the time derivative of the radial velocity is highest, and weakest at quadrature (solid line), where the derivative is smallest. Two intermediate phases are also shown. The FWHM given by Eq.~\ref{eq:doppler} is indicated by a horizontal red line; orbital broadening becomes significant as the curves approach this line. In general, observations made at quadrature with exposure times up to ${\sim} 6$ hours cause negligible broadening. At all other phases, however, broadening becomes significant in a matter of one or a few hours. At full and new phase, the line width doubles after an exposure of only 40 minutes. However, at these phases the radial velocity of the planet relative to the star is zero, and as we argue in \S\ref{sec:search} below, disentangling stellar and planetary emission becomes difficult. In the discussion that follows, we therefore focus on observations made close to quadrature.

\begin{figure}[bt]
\includegraphics[width=0.47\textwidth]{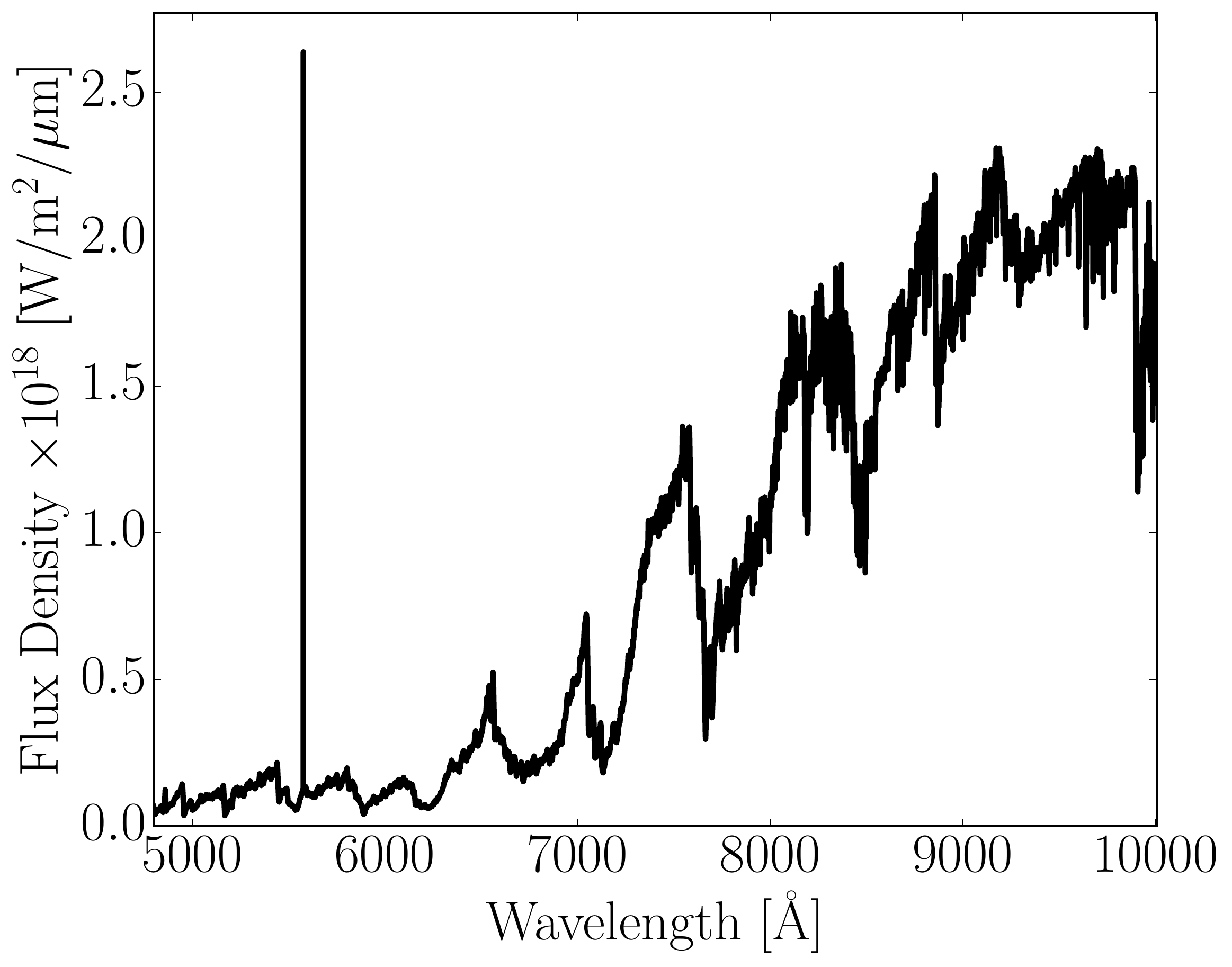}
\caption{Simulated high-resolution visible spectrum of Proxima Cen b with a 0.1 TW OI auroral emission at $5577$ \AA. A grey geometric albedo of 0.3 is assumed for the planet. The spectrum is calculated at quadrature phase and scaled to the observing distance (1.302 pc).}
\label{fig:spec}
\end{figure}

Fig.~\ref{fig:spec} shows a high-resolution model spectrum of Proxima Cen b at quadrature, illustrating an auroral emission feature that could be expected from the planet. We injected a Gaussian line at 5577\AA\ with $\text{FWHM} = 0.014$\AA, normalized to a steady-state Proxima Cen b OI auroral power of $L_{OI} = $ 0.1 TW. This OI auroral power yields an equivalent width of ${\sim} 3.63 (L_{OI}/1 {\rm \ TW})$\AA\ relative to our model of the reflected planetary spectrum at quadrature.

\subsection{Contrast Ratios \& Telescope Integration Times}
\label{sec:int_times}

To unambiguously detect a narrow emission feature, such as the example shown in Fig.\ \ref{fig:spec}, the telescope resolving power ($R \equiv \lambda / \Delta \lambda$) needs to be taken into account. The typical resolving power being considered for future space-based coronagraph mission concepts is $R \approx 100$, which is appropriate for the detection of molecular absorption bands in the optical and NIR given the relatively low planet-star contrast ratio \citep{Robinson2016}. However, at 5577\AA\ an $R=100$ spectrograph has a spectral element width of $\Delta \lambda \approx 56$\AA, over ${\sim} 10^3$ times broader than the OI green line width. Future space-based high-contrast exoplanet imaging missions would need to fly with higher resolution spectrographs to detect the OI $5577$\AA\ line.

\begin{figure}[bt]
\includegraphics[width=0.49\textwidth]{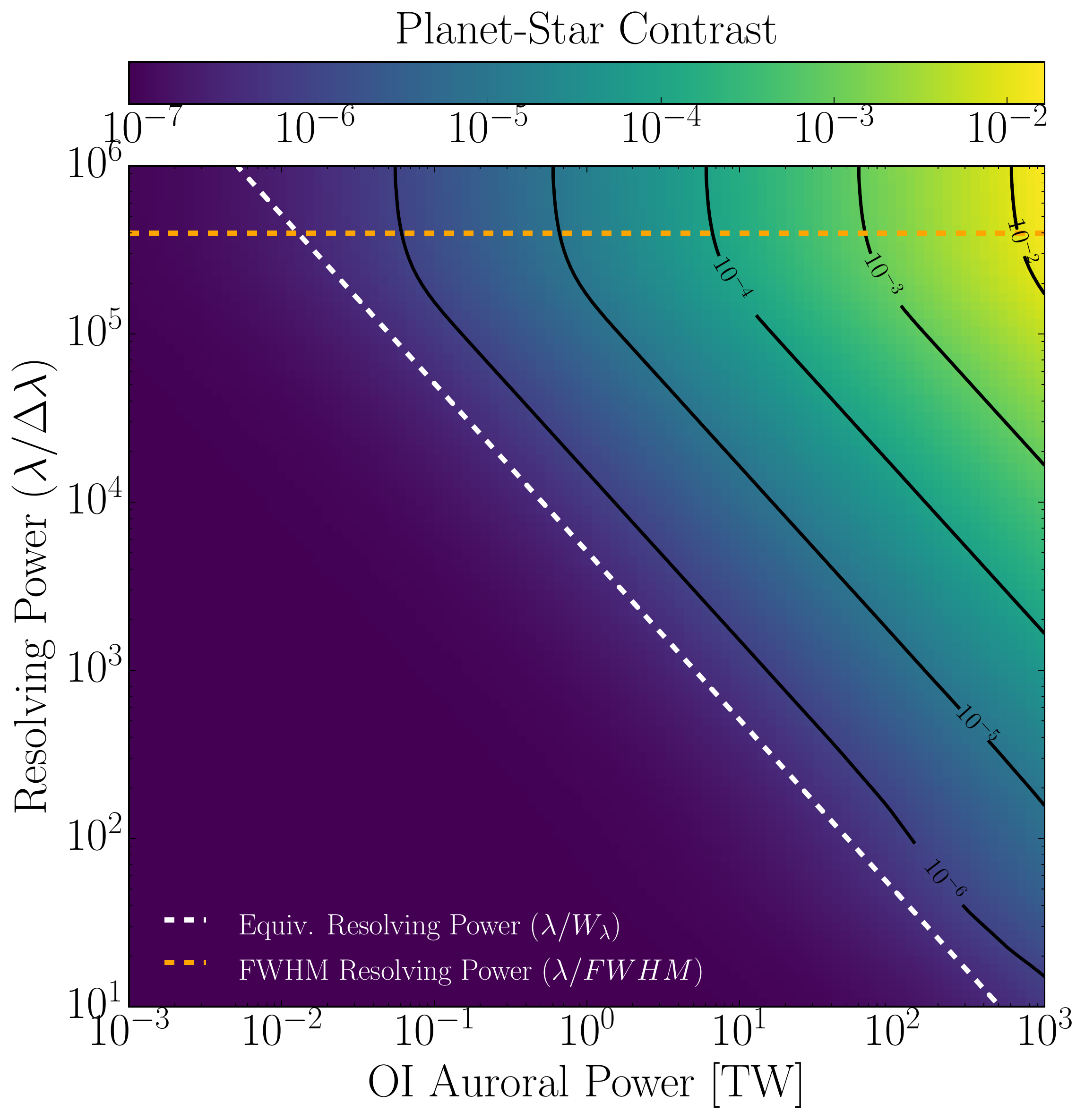}
\caption{Planet-star contrast ratio contours as a function of telescope resolving power and OI auroral power. The full width at half maximum (FWHM; dashed-orange) of the line and equivalent width ($W_{\lambda}$) as a function of auroral power (dashed-white) are represented as ``resolving powers'', where $R_{FWHM} = \lambda_{\text{OI}} / FWHM$ and $R_{W} = \lambda_{\text{OI}} / W_{\lambda}$, respectively. The black contour lines show curves of constant planet-star contrast.}
\label{fig:contrast}
\end{figure}

Fig.~\ref{fig:contrast} shows planet-star contrast ratios in a spectral element centered on the 5577\AA\ OI auroral line as a function of spectrograph resolving power and auroral power, assuming a FWHM of 0.014\AA\ (i.e., negligible orbital broadening). The FWHM of the auroral line and equivalent width ($W_{\lambda}$) as a function of auroral power are represented as ``resolving powers,'' where $R_{FWHM} = \lambda_{\text{OI}} / FWHM$ and $R_{W} = \lambda_{\text{OI}} / W_{\lambda}$, respectively. In Fig.~\ref{fig:contrast}, the dashed-white line gives the resolving power such that the spectral element width matches the equivalent width of the line at a given auroral power.  The dashed-orange line gives the resolving power such that the spectral element width matches the FWHM of the line. That is, a fixed FWHM=0.014\AA\ yields $R_{FWHM} = 4 \times 10^5$. Optimal observations occur when the planet-star contrast (indicated by the contours) is highest. An increase in the contrast of the emission line is only achieved when the width of a spectral element is smaller than the equivalent width of the line. For resolving powers greater than $R_{FWHM}$, multiple spectral elements are needed to span the emission line, which may introduce additional unwanted read noise and dark current. Therefore, observations should be made in the wedge between the FWHM resolving power and the equivalent width resolving power.  Our predicted steady-state auroral emission (${\sim}0.1$~TW) requires that spectrographs achieve $R \gtrsim 10^5$.

\begin{deluxetable*}{cccccccccc}[th!]
\tablewidth{\linewidth}
\tablenum{4}
\tablecaption{Planet-Star contrast ratios and telescope integration times necessary to detect the $5577$\AA\ OI auroral line}
\tablehead{
\colhead{} & \colhead{} & \multicolumn{8}{c}{Telescope Integration Time [hours]} \\
\cline{3-10} \\
\colhead{Power [TW]} & \colhead{Contrast} & \colhead{HARPS} & \colhead{VLT} & \colhead{VLT + C} & \colhead{TMT} & \colhead{TMT + C} & \colhead{HabEx} & \colhead{LUVOIR} & \colhead{TMT + C*}
}
\startdata
$\phantom{000}0.001$  & $9 \times 10^{-8}$ & $2 \times 10^{13}$ & $4 \times 10^{12}$ & $4 \times 10^{8}$  & $3 \times 10^{11}$  & $1 \times 10^{7}$  & $6 \times 10^{8}$  & $2 \times 10^{7}$           & $1 \times 10^{5}$  \\
$\phantom{000}0.01\phantom{0}$ & $2 \times 10^{-7}$ & $2 \times 10^{11}$ & $4 \times 10^{10}$ & $4 \times 10^{6}$  & $3 \times 10^{9}$   & $1 \times 10^{5}$  & $6 \times 10^{6}$  & $2 \times 10^{5}$  & $2 \times 10^{3}$  \\
$\phantom{000}\mathbf{0.1}\phantom{00}$ & $\mathbf{8 \times 10^{-7}}$ & $\mathbf{2 \times 10^{9}}$  & $\mathbf{4 \times 10^{8}}$  & $\mathbf{4 \times 10^{4}}$  & $\mathbf{3 \times 10^{7}}$   & $\mathbf{1 \times 10^{3}}$  & $\mathbf{6 \times 10^{4}}$  & $\mathbf{2 \times 10^{3}}$  & $\mathbf{40}$               \\
$\phantom{000}1\phantom{.000}$ & $7 \times 10^{-6}$ & $2 \times 10^{7}$  & $4 \times 10^{6}$  & $4 \times 10^{2}$  & $3 \times 10^{5}$   & $10$               & $7 \times 10^{2}$  & $30$               & $3$                \\
$\phantom{00}10\phantom{.000}$ & $7 \times 10^{-5}$ & $2 \times 10^{5}$  & $4 \times 10^{4}$  & $7$                & $3 \times 10^{3}$   & $4 \times 10^{-1}$ & $10$               & $1$                & $3 \times 10^{-1}$ \\
$\phantom{0}100\phantom{.000}$ & $7 \times 10^{-4}$ & $2 \times 10^{3}$  & $4 \times 10^{2}$  & $4 \times 10^{-1}$ & $30$                & $3 \times 10^{-2}$ & $6 \times 10^{-1}$ & $9 \times 10^{-2}$ & $3 \times 10^{-2}$ \\
$1000\phantom{.000}$ & $7 \times 10^{-3}$ & $20$               & $4$                & $3 \times 10^{-2}$ & $3 \times 10^{-1}$ & $3 \times 10^{-3}$ & $6 \times 10^{-2}$ & $9 \times 10^{-3}$            & $3 \times 10^{-3}$
\tablecomments{\label{tab:OI_detect} Integration times refer to the time required to achieve a signal-to-noise of 6 on the auroral emission above the continuum assuming a telescope throughput of $5\%$, a spectrograph with resolution $\lambda / \Delta \lambda = 115,000$ for HARPS, TMT, HabEx and LUVOIR and $\lambda / \Delta \lambda = 120,000$ for VLT.  ``+ C'' indicates the use of a coronagraph and associated noise sources discussed in \citet{Robinson2016}. Auroral power of order $0.1$~TW (boldface) corresponds to the predicted steady-state value in \S\ref{sec:signal} while ${\sim}1 - 100$~TW corresponds to our predicted auroral power arising from a combination of substorm event and CMEs.  TMT + C$^{*}$ denotes a coronagraph-equipped TMT concept with a design contrast of $C = 10^{-7}$ and negligble instrumental noise.}
\end{deluxetable*}

Using the auroral power estimates from \S\ref{sec:signal}, we explore the feasibility of detecting the 5577\AA\ OI auroral emission line with five different ground-based telescope configurations: the 3.6m High Accuracy Radial velocity Planet Searcher (HARPS), the 8.2m Very Large Telescope (VLT) with and without a coronagraph, and a Thirty Meter Telescope (TMT) concept with and without a coronagraph \citep{Skidmore2015,Udry2014,Johns2012}. We also model the detection using two future space-based coronagraph concepts: the 16m Large UV/Optical/IR Surveyor \citep[LUVOIR;][]{Kouveliotou2014,Dalcanton2015} and the 6.5m Habitable Exoplanet Imaging Mission \citep[HabEx;][]{Mennesson2016}.

High spectral resolution coronagraphy with the VLT will require an update to the SPHERE high-contrast imager and a coupling with the ESPRESSO spectrograph, as described in \citet{Lovis2016}. Note that given the VLT's 8.2m diameter, the SPHERE coronagraph must achieve an inner working angle no more than $\theta_{\text{IWA}} = 2.7 \lambda / D$ to observe at wavelengths as long as 5577\AA\, given the maximal planet-star angular separation of 37 mas for Proxima Cen b. Our HARPS, TMT, LUVOIR and HabEx telescope models use $R = 115,000$, while for VLT we use $R = 120,000$. All models assume a total telescope and instrument throughput of $5\%$ and a quantum efficiency of 90\%. Coronagraph noise estimates use the model presented in \citet{Robinson2016} with updated parameters from \citet{Meadows2016}, and consider noise due to speckles, dark counts, read noise, telescope thermal emission, and zodi and exozodi light. Ground-based coronagraphy assumes a conservative design contrast of $10^{-5}$ \citep{Dou2010,Guyon2012}, while space-based assumes $10^{-10}$ \citep{Meadows2016} unless stated otherwise. Typically, telescope detectors have a maximum exposure time to mitigate the damaging effect of cosmic ray strikes \citep[see][]{Robinson2016}. Therefore, integration times longer than one hour require multiple readouts, introducing more detector noise. Non-coronagraph telescope calculations assume only stellar noise at the photon limit; their values are therefore lower limits, and may increase significantly due to stellar activity (see \S\ref{sec:search}). To prevent significant orbital broadening, we assume that observations are made for one hour at a time at or close to quadrature; longer exposure times are achieved by stacking multiple observations. For exposure times much longer than an hour, stacking will appreciably increase the read noise and dark current for coronagraph observations, where the star is nulled, but not for the non-coronagraph observations, where the stellar photons dominate the noise budget.

Our integration time calculations follow those described in \citet{Robinson2016}. For the stellar spectrum we adopt the steady-state Proxima Centauri spectrum of \citet{Meadows2016} and neglect the impact of flares on the stellar continuum. We assume that the quoted auroral power emitted via the 5577\AA\ OI line is constant throughout the entire observation.

For observations without a coronagraph, both the stellar flux and reflected stellar flux define the continuum from which we wish to resolve the auroral emission feature. Observations with a coronagraph need only resolve the auroral emission above the coronagraph noise and reflected stellar continuum.  With these considerations in mind, we simulate the net planetary emission as a combination of reflected stellar continuum and auroral emission.  We compute the flux from the reflected stellar continuum by assuming that the planet is a Lambertian scatterer at quadrature with a planetary geometric albedo of 0.3 and a planetary radius of $1.07\mathrm{R}_{\oplus}$  following \citet{Barnes2016}.  We then inject the expected flux from the auroral line at its wavelength.  We integrate over all spectral elements that contain the auroral line flux, taking the auroral photon count rate as our signal and all other sources as noise as in \citet{Robinson2016}.  For the oxygen 5577\AA\ line width of ${\sim} 0.014$\AA\ (\S~\ref{sec:line_profile}) and our nominal resolving power, this corresponds to one spectral element.

Table~\ref{tab:OI_detect} shows the integration times required to achieve a signal-to-noise of 6 on the 5577\AA\ OI auroral emission line above the stellar and reflected planetary continuum as a function of auroral power.  We simulated contrast ratios and integration times for emitted auroral powers at the OI 5577\AA\ line ranging from $10^{-3} - 10^{3}$~TW to bracket all potential auroral fluxes. The $10^{-3}$~TW lower limit corresponds to a strong 5577\AA\ emission from Earth (Earth total electromagnetic auroral power is of order $10^{-2}$~TW with 5577\AA\ typically $\sim$2\% of this value). The upper limit of $10^{3}$~TW is an extreme case that is an order of magnitude stronger than the largest value predicted in \S\ref{sec:signal}. Values in between correspond to the different cases considered in Table~\ref{tab:auroral_power}, which depend on the planetary dipole moment, magnetospheric substorm activity, and whether CME conditions are present. For reference, the estimated steady-state Proxima Cen b value calculated in \S\ref{sec:signal} is ${\sim}0.1$~TW.

The weak $10^{-3}$~TW aurora is indistinguishable from the purely reflecting planet-star contrast near the 5577\AA\ OI auroral emission line \citep{Turbet2016,Meadows2016} and effectively demonstrates why high resolution spectroscopy is not typically considered for high-contrast imaging.
For an Earth-like planet, the auroral power estimates from \S\ref{sec:signal} (${\sim} 0.1$~TW) make detecting the OI emission line infeasible with current instruments, even though the contrast ratio at the line is relatively strong (${\sim}8 \times 10^{-7}$). 
Although unlikely, if the auroral power were much higher (${\sim} 10^{3}$~TW) and sustained over the period of the observation, detection of OI emission could be possible with current instruments in tens of hours.
Realistically, however, these estimates suggest that current instruments are likely not capable of detecting an OI aurora on Proxima Cen b.

For a SPHERE-ESPRESSO coupling \citep{Lovis2016}, the integration times required to detect an OI auroral line are slightly more favorable over a wide range of possible auroral powers, but still prohibitively long under most plausible circumstances. If Proxima Cen b has a Neptune-strength magnetic dipole moment, and observations were made during substorm conditions (when the power in the OI 5577\AA\ line reaches ${\sim} 1$~TW with a contrast ratio of $7 \times 10^{-6}$), a coronagraph-equipped VLT would have to integrate for ${\sim} 400$ hours. However, if observations coincided with periods of more vigorous stellar activity such as during a concurrent CME and substorm, the auroral output could reach ${\sim} 100$~TW and contrast ratios of $7 \times 10^{-4}$.  An upgraded SPHERE (denoted by VLT+C in Table~\ref{tab:OI_detect}) may be able to detect this signal in under an hour. Since CMEs and fast stellar wind streams can have timescales ${\sim} 10$ hours, and substorms up to several days \citep{Gonzalez1994,Gonzalez1999}, the high level of transient activity may be observable.  Under near constant CME activity, storm conditions could potentially last for weeks or longer, improving the chances of detecting auroral emission  \citep{Khodachenko2007}.

Even future observations with TMT, HabEx, and LUVOIR outfitted with instruments optimized for high-resolution, high-contrast coronagraphy will be unable to detect a steady-state 0.1~TW OI aurora on Proxima Cen b. However, a coronagraph-equipped TMT could detect a substorm strength aurora of ${\sim} 10$ TW in about 10 hours, while LUVOIR could make the predicted substorm auroral observation in about 30 hours. Only auroral powers ${\gtrsim}10$ TW would be detectable with HabEx.
Auroral powers of order $100$~TW arising from a concurrent CME and substorm could be observed by the TMT, HabEx, and LUVOIR in well under an hour.

\begin{figure}
\centering
\includegraphics[width=0.47\textwidth]{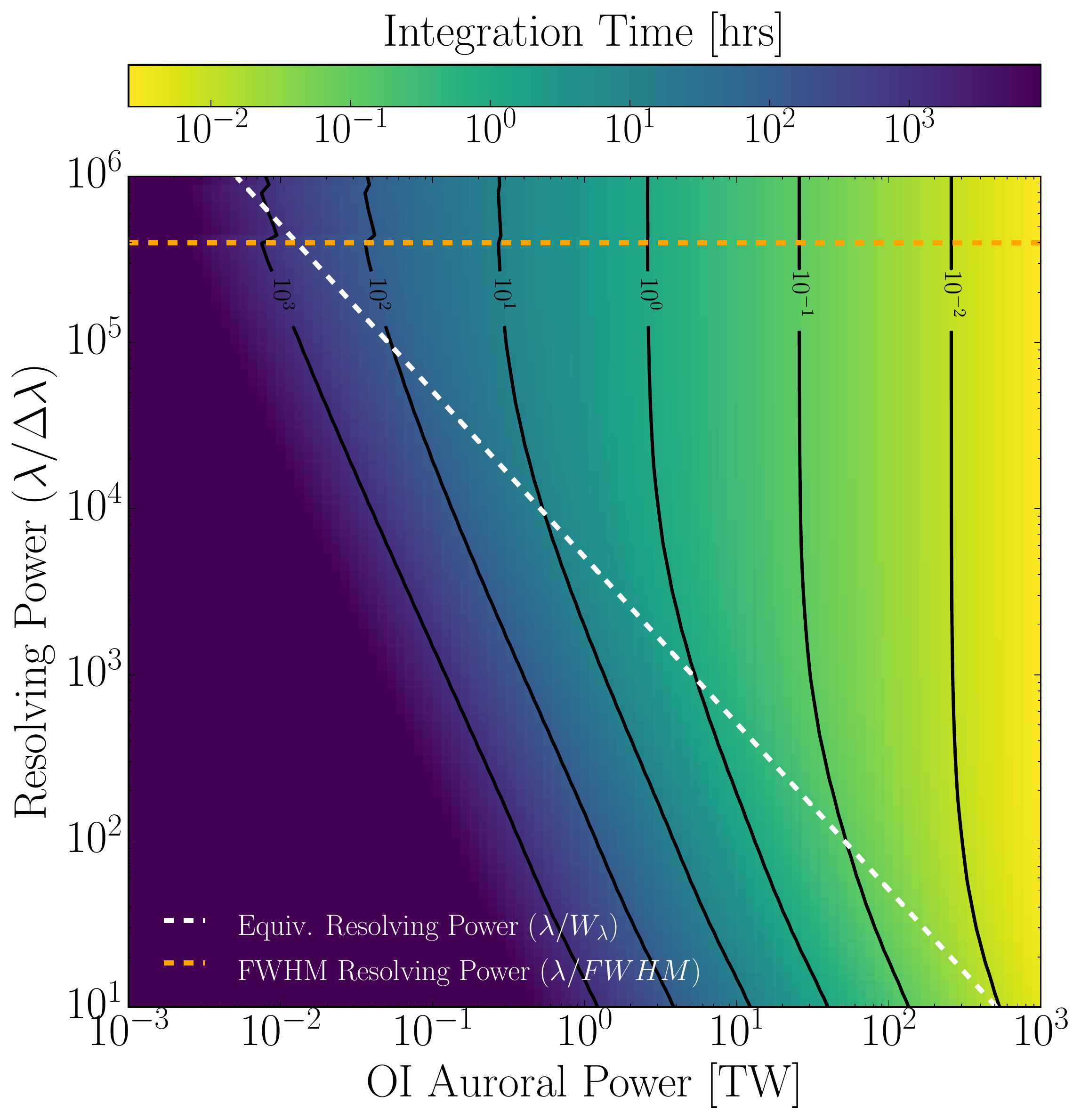}
\includegraphics[width=0.47\textwidth]{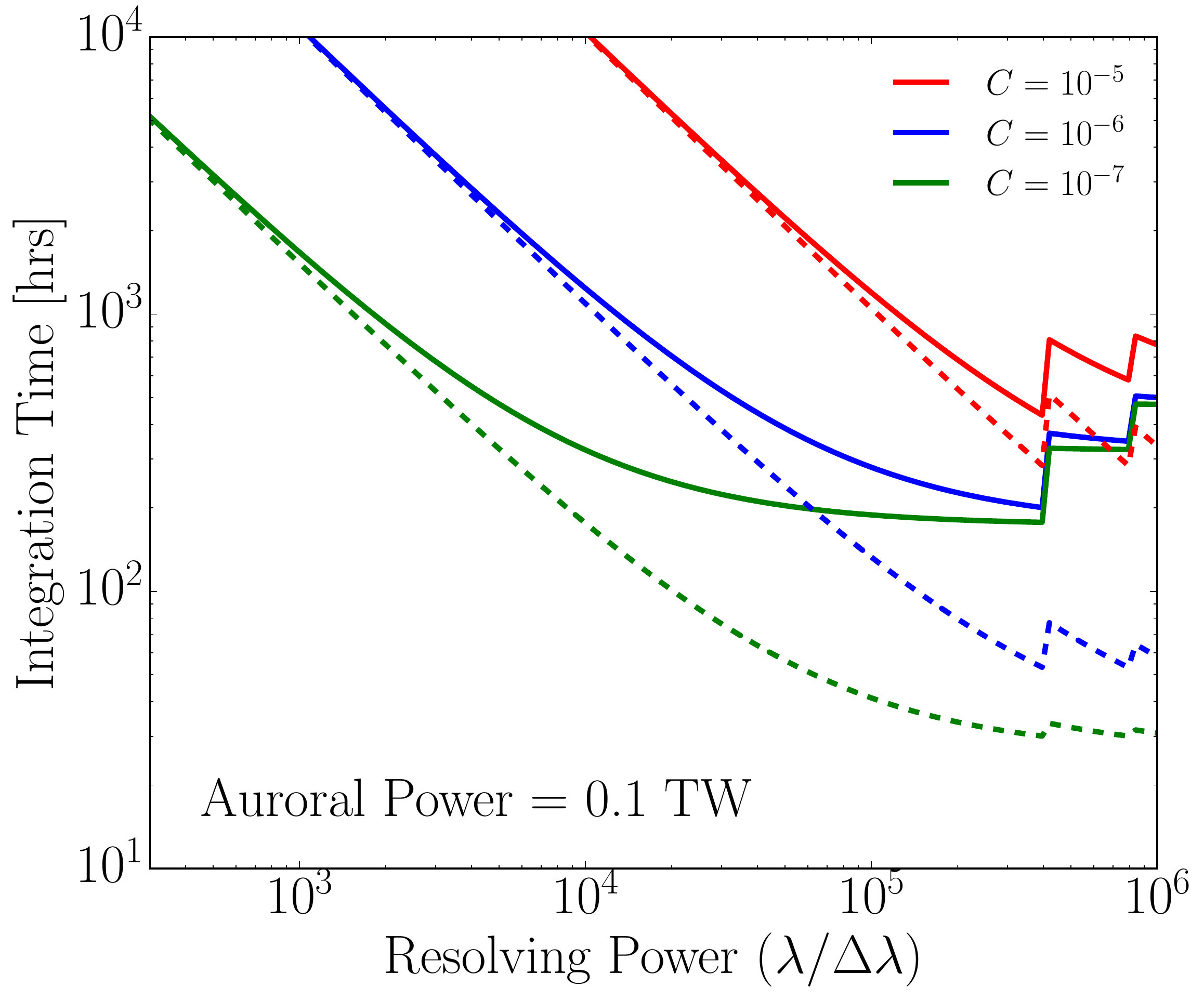}
\caption{\textit{Top}: Similar to Fig.~\ref{fig:contrast}, but displays telescope integration time contours as a function of telescope resolving power and OI auroral power for a coronagraph-equipped TMT concept with a design contrast of $C = 10^{-7}$. Dark current, read noise, and telescope thermal noise are set to zero here to simulate optimal detector performance that may be achieved by future instruments. \textit{Bottom}: Telescope integration time as a function of resolving power for a coronagraph-equipped TMT concept for three different design contrasts.  The solid curves denote integration times that include all modeled noise sources while the dashed curves assume negligible instrumental noise.}
\label{fig:TMTC}
\end{figure}

Finally, we consider how improvements in ground-based instrumentation might expand the ability to detect exo-aurorae. In the top panel of Fig.~\ref{fig:TMTC}, we model a coronagraph-equipped TMT concept that achieves a design contrast of $C = 10^{-7}$ and has negligible instrumental noise (e.g., no dark current and read noise).  Low-resolution observations with a resolving power smaller than the equivalent width resolving power, $R < \lambda/W_\lambda$, yield longer integration times at fixed auroral power as the auroral signal is diluted by additional stellar continuum photons from larger spectral elements, which increases the noise.  For high resolutions that exceed the equivalent width resolving power, $R > \lambda/W_\lambda$, the auroral signal dominates the planetary continuum as the spectral element more tightly bounds the narrow emission feature, yielding little additional improvement in integration times.

In the bottom panel of Fig.~\ref{fig:TMTC}, we vary coronagraph design contrasts for observations with and without instrumental noise.  We find that a TMT with coronagraphic starlight suppression, negligible instrumental noise, a design contrast of $C = 10^{-7}$ and $R > 10^5$ allows for a detection of our predicted steady-state OI auroral emission (auroral power of ${\sim}0.1$~TW) over about 40 hours (see also Table~\ref{tab:OI_detect}). The discontinuities that occur at high resolving powers are due to the need for additional spectral elements to span the width of the OI auroral line.

Despite the likely increased strength of aurorae on Proxima Cen b compared to Earth, observing a {\em steady-state} 0.1~TW aurora requires sufficiently long integration times that it is not currently feasible, nor will it be feasible with the next generation of instruments, unless ideal instrumental performance were achieved. OI auroral detection may only be possible if observations coincide with magnetospheric substorms or periods of vigorous stellar activity, such as CMEs, which can induce much stronger aurorae ranging from $1 - 10$~TW (and up to ${\sim} 100$~TW if Proxima Cen b has a stronger magnetic dipole than Earth). These transient events are frequent on Proxima Centauri \citep{Davenport2016} and may persist on timescales comparable to the integration times needed to detect strong aurorae.

\section{Search in the HARPS Data}
\label{sec:search}

The ESO Archive\footnote{\url{http://archive.eso.org/}} hosts 319 HARPS spectra of Proxima Centauri taken between 2004 and 2016 and totaling about 70 hours of exposure time. The spectra were taken in the wavelength range $3782-6913$\AA\ with a resolving power $R = 115,000$, yielding a wavelength resolution $\Delta\lambda \approx 0.05$\AA\ at 5577\AA. Each wavelength bin was oversampled by a factor of about 5. Given the estimates in Table~\ref{tab:OI_detect}, if Proxima Cen b's auroral power were on the order of 10$^{3}$~TW (however unlikely), the OI line could be detectable in this dataset. We therefore downloaded all spectra to conduct a search for the OI emission feature of Proxima Cen b. The method we outline below is similar to so-called ``spectral deconvolution'' techniques used to detect molecular absportion in exoplanet atmospheres \citep[e.g.,][]{SparksFord2002,RiaudSchneider2007,Kawahara2014,Snellen2015}.

We first shifted all spectra to the stellar rest frame by cross-correlating them against each other and calibrating the wavelength array to the stellar Na D I and II lines. Next, we removed stellar lines by performing weighted principal component analysis \citep[WPCA;][]{Delchambre2015} on a 250\AA\ window centered at 5577\AA. Each spectrum was then fit with a linear combination of the first 10 principal components, a number which we obtained by optimizing the recovery efficiency of injected planetary signals (see below); the fit was then subtracted, reducing the noise in the vicinity of 5577\AA\ by a factor of ${\sim} 7$. In order to obtain the principal components, we weighted each spectrum by the square root of its exposure time and assigned weights of zero to the individual telluric 5577\AA\ airglow features, as these are among the strongest features in any individual spectrum and may incorrectly bias the principal components in the stellar frame. We remove Earth airglow separately below.

\begin{figure}[bt]
\includegraphics[width=0.47\textwidth]{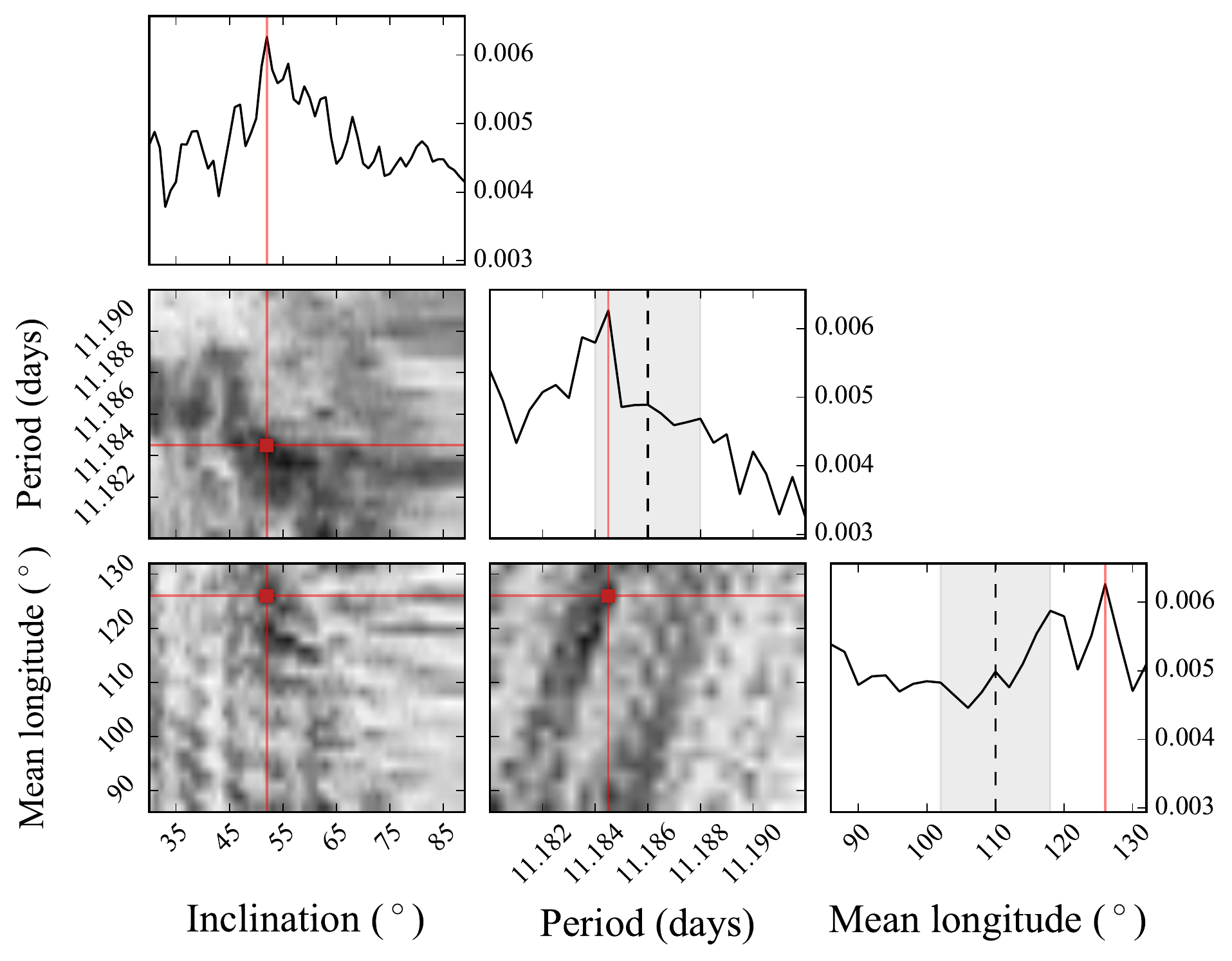}
\caption{Results from the grid search over inclination ($i$), period ($P$), and mean longitude ($\lambda$) for the strongest 5577\AA\ planetary signal. The inclination grid spans the range $30^\circ - 90^\circ$ in increments of $1^\circ$. The period and mean longitude grids are centered on the best-fit values reported in \citet{Anglada-Escude2016} and span the $\pm 3\sigma$ range in increments of $0.25\sigma$. In total, 37,440 different orbital configurations for Proxima Cen b were considered. The curves along the main diagonal show the fractional amplitude of the bin centered on the OI line as a function of inclination (top left), period (center), and mean longitude (bottom right). In the period and mean longitude plots, the dashed line is the value reported in the discovery paper, with the 1$\sigma$ bounds shaded in gray. The colormaps show the joint distributions of signal strengths for pairs of the three orbital parameters (black highest, white lowest). The peak signal is indicated by the red lines and occurs at $i = 52^\circ$, $P = 11.1845$ days, and $\lambda = 126^\circ$, with detection significance ${\sim}0.7\sigma$. As we argue below, this signal has a very high false alarm probability (FAP ${\sim} 0.2$) and is entirely consistent with noise.}
\label{fig:triangle}
\end{figure}

\begin{figure}[bt]
\includegraphics[width=0.47\textwidth]{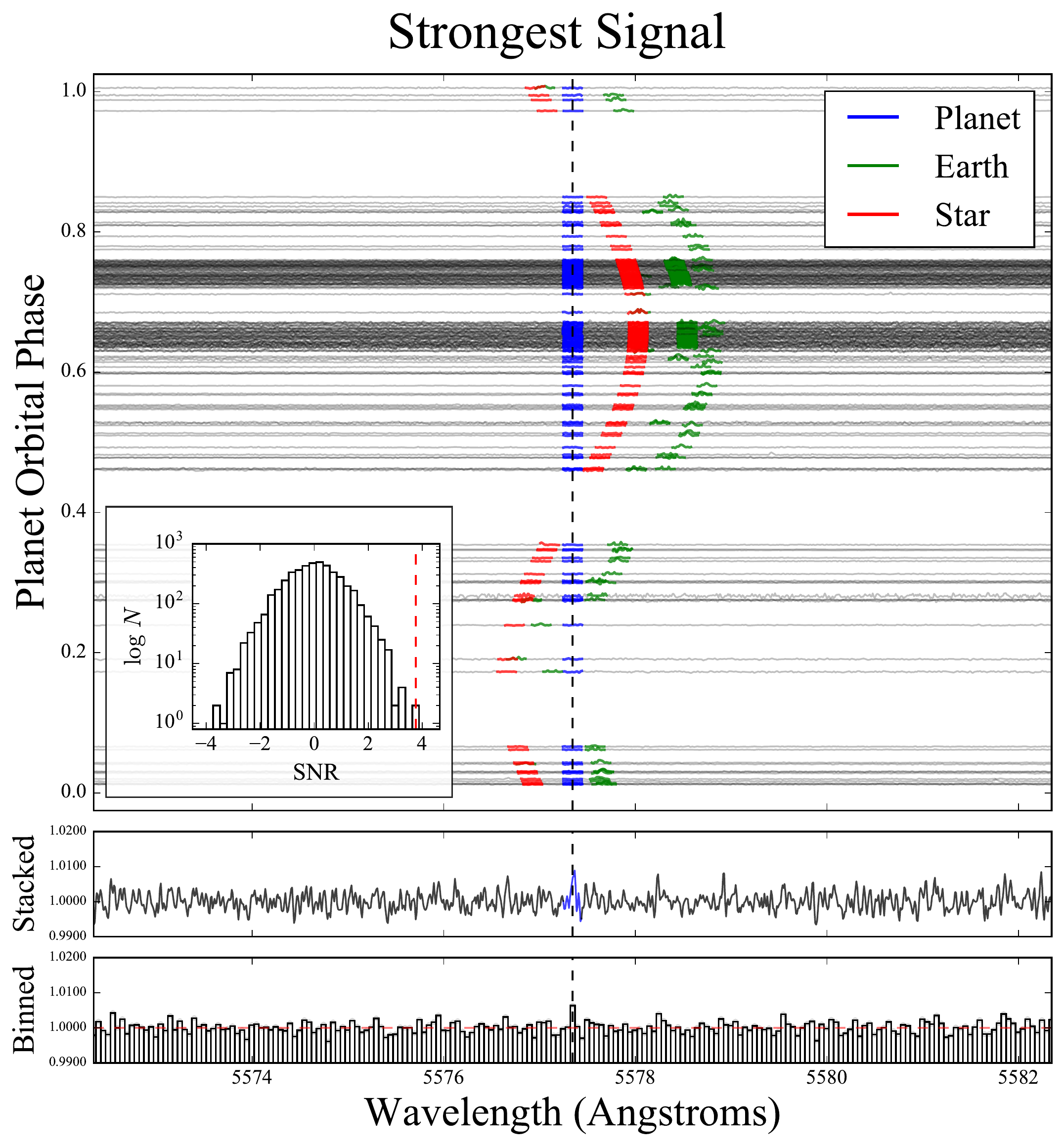}
\caption{The HARPS spectra of Proxima Centauri. After removing stellar and telluric lines, the individual spectra are Doppler-shifted into the frame of Proxima Cen b according to the orbital parameters corresponding to the peak signal in Fig.~\ref{fig:triangle}. The spectra are then normalized and distributed vertically on the main subplot according to the planet's orbital phase. Blue regions indicate a small (0.2\AA) window centered on the 5577\AA\ oxygen feature in the planet frame. Red and green regions indicate the same window in the star and Earth frames, respectively; note the residual telluric airglow features in many of the spectra. The bottom subplots show the stacked spectrum in the planet frame and the stacked spectrum after downsampling to bins of size equal to the instrumental FWHM of the line (0.05\AA). The peak recovered by the grid search is evident in both the stacked and the binned flux. The inset at the center left shows a histogram of the amplitude of deviations from the median in bins across a 250\AA\ window centered on the OI line, indicating a signal-to-noise ratio (SNR) of about 4 in the 5577\AA\ bin. Despite the apparent strength of this detection, further analysis yields a detection significance of only ${\sim} 0.7\sigma$, with false alarm probability ${\sim} 20\%$ (see Fig.~\ref{fig:fap}).}
\label{fig:strongest_river}
\end{figure}

Next, we Doppler-shifted all spectra into the frame of Proxima Cen b. Since the orbital inclination $i$ is unconstrained, we performed a grid search, varying $i$ in one degree increments from 30$^\circ$ to 90$^\circ$. We did not consider inclinations lower than 30$^\circ$ due to the difficulty of deconvolving stellar and planetary signals in near face-on orbits. We further varied the planet period $P$ and planet mean longitude $\lambda$ across a range spanning $\pm 3\sigma$ about the best fit values reported in Table~1 of \citet{Anglada-Escude2016}, in increments of $0.25\sigma$; in total, we considered 37,440 different orbital configurations for the planet. For simplicity, the eccentricity was assumed to be zero, the planet mass was set to $1.27\mathrm{M_\oplus}/\sin i$, and the stellar mass was held fixed at 0.12$\mathrm{M_\odot}$. The latter parameter is considerably uncertain; however, changing the stellar mass changes the amplitude of the planetary RV signal, making the stellar mass degenerate with the inclination of Proxima Cen b's orbit. A grid search over the stellar mass is therefore redundant as long as we treat the inclination above as an ``effective'' inclination for $M_\star = 0.12\mathrm{M_\odot}$.

After Doppler-shifting each spectrum, we translated them back to the original wavelength grid by linear interpolation. Once in the planet frame, we identified and interpolated over $> 10 \sigma$ outliers in each wavelength bin of the normalized spectra outside the 0.2\AA\ window centered on the OI line. We found that this successfully removed telluric airglow and prevented outlier features in individual spectra from contributing to the stacked spectrum. We purposefully did not perform this outlier removal step in the vicinity of the (putative) planetary 5577\AA\ line to prevent time-variable emission from being removed. Note that, in principle, this could result in a false detection of a planetary signal due to the presence of a large (non-planetary) outlier in a single spectrum. In the event that a signal were recovered, a detailed analysis of the spectrum/spectra it originated from would be necessary to rule out this possibility.

For each orbital configuration, we then co-added all spectra in the planet frame, omitting spectra in which the planetary 5577\AA\ window overlapped with either the stellar or telluric 5577\AA\ windows to avoid contamination from OI emission by those sources. For orbits close to edge-on, this reduced the total exposure time from 70 to about 50 hours, and less for lower inclination orbits. In order to remove correlated stellar noise, we then applied a high pass median filter of window size 1\AA.

Finally, we binned the stacked spectra to 0.05\AA-wide bins, with the central bin centered at 5577.345\AA, the empirical wavelength of the OI green line \citep{CabannesDufay1955,Chamberlain1961}. Our bin size is the HARPS resolution at that wavelength, and closely matches the FWHM of the telluric OI lines in the dataset. As we argued in \S\ref{sec:detect}, a higher resolution spectrograph (with less instrumental broadening) would allow for smaller bin sizes and higher contrast in the OI line. We then measured the amplitude of the 5577.345\AA\ bin relative to the spectrum mean.

The results of our grid search are shown in the triangle plot in Fig.~\ref{fig:triangle}. Along the main diagonal, we plot the maximum fractional strength of the 5577\AA\ signal as a function of each of the orbital parameters (the inclination $i$, the period $P$, and the mean longitude $\lambda$). Below those plots, we show the two-parameter joint distributions of the maximum signal strength, where darker colors correspond to higher values. A peak is visible at an (effective) inclination of 52$^\circ$, a period of 11.1845 days, and a mean longitude of 126$^\circ$. In Fig.~\ref{fig:strongest_river} we show the spectra Doppler-shifted into the planet frame according to these orbital parameters. Each of the processed spectra are normalized and distributed vertically along the main subplot according to the planetary phase at the time the observation was made. The location of the expected OI planetary feature is indicated in blue; we show the same window in the frame of Proxima Centauri (red) and Earth (green), where residual telluric emission is clearly visible. As mentioned above, spectra in which Proxima Cen b's 5577\AA\ window overlaps with either the stellar or telluric windows are omitted. When stacking the spectra below, we also we masked and interpolated over 0.2\AA\ windows centered on the telluric features.

\begin{figure}[bt]
\includegraphics[width=0.47\textwidth]{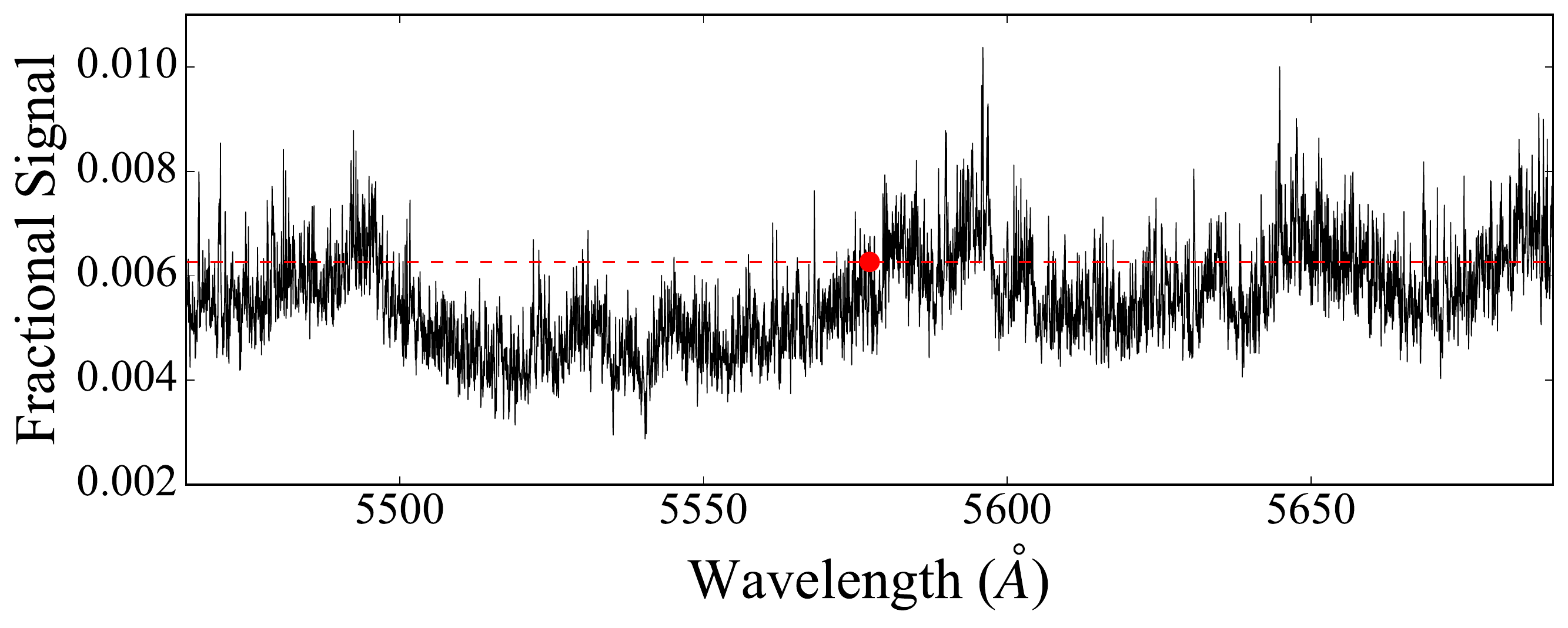}
\caption{The peak signal in each wavelength bin in the vicinity of the 5577\AA\ line. The fractional signal ($y$ axis) is the flux in the bin divided by the continuum, and would roughly correspond to a planet-star contrast ratio if the signal were real. The peak signal at the 5577\AA\ line ($0.7\sigma$) is indicated by the dashed red line. About 20\% of the bins display stronger peak signals than the 5577\AA\ bin, leading to a FAP for the 5577\AA\ signal of ${\sim} 20\%$. Note also the strong correlated noise as a function of wavelength, likely due to improperly subtracted time-variable stellar features.\\[0in]}
\label{fig:max_signal_vs_wavelength}
\end{figure}

\begin{figure}[bt]
\includegraphics[width=0.47\textwidth]{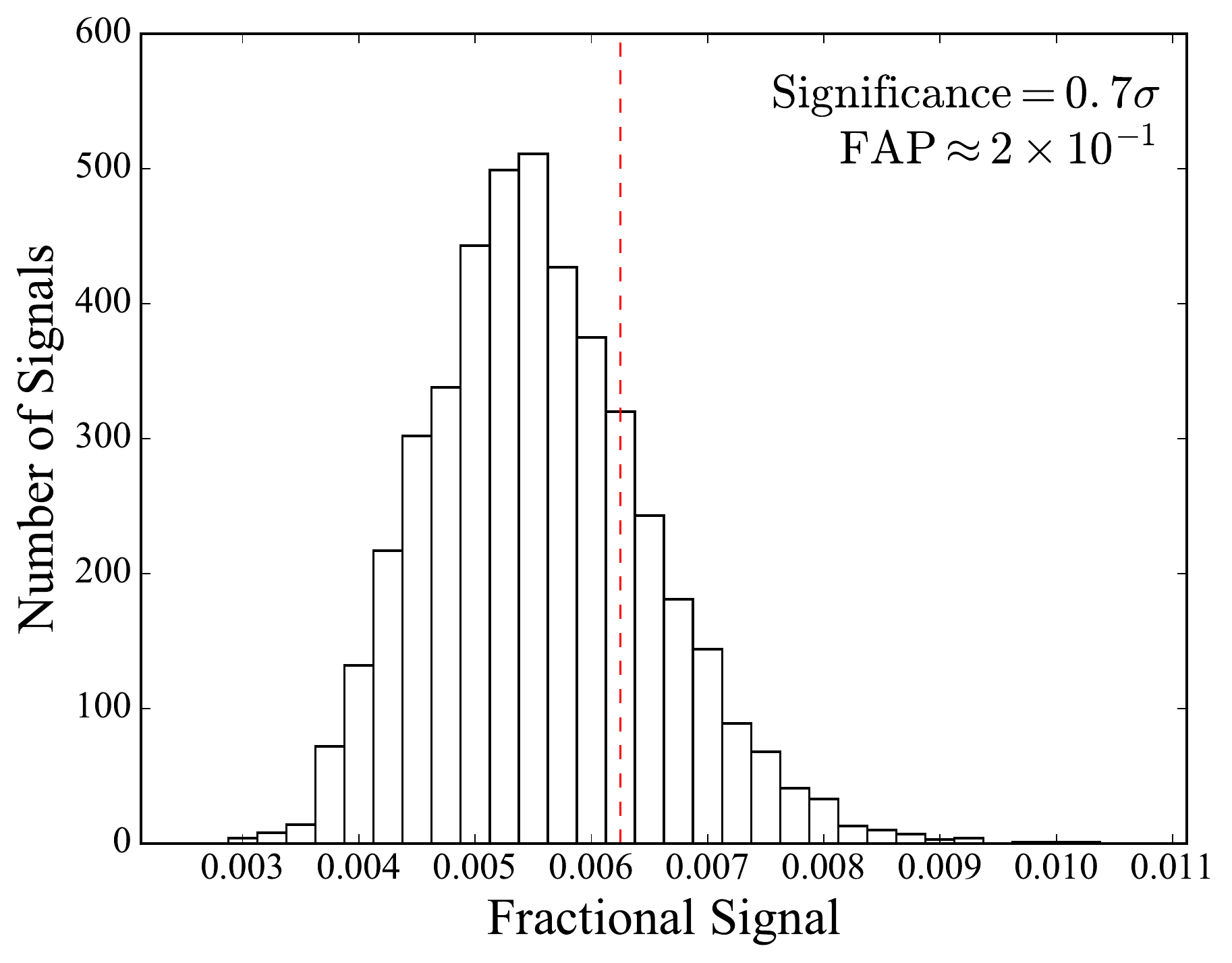}
\caption{The distribution of the signal strength over the wavelength grid in Fig.~\ref{fig:max_signal_vs_wavelength}. The bin corresponding to the peak 5577\AA\ signal is indicated with a red dashed line; given the large FAP, the recovered signal is fully consistent with noise.\\[0in]}
\label{fig:fap}
\end{figure}

\begin{figure}[bt]
\includegraphics[width=0.47\textwidth]{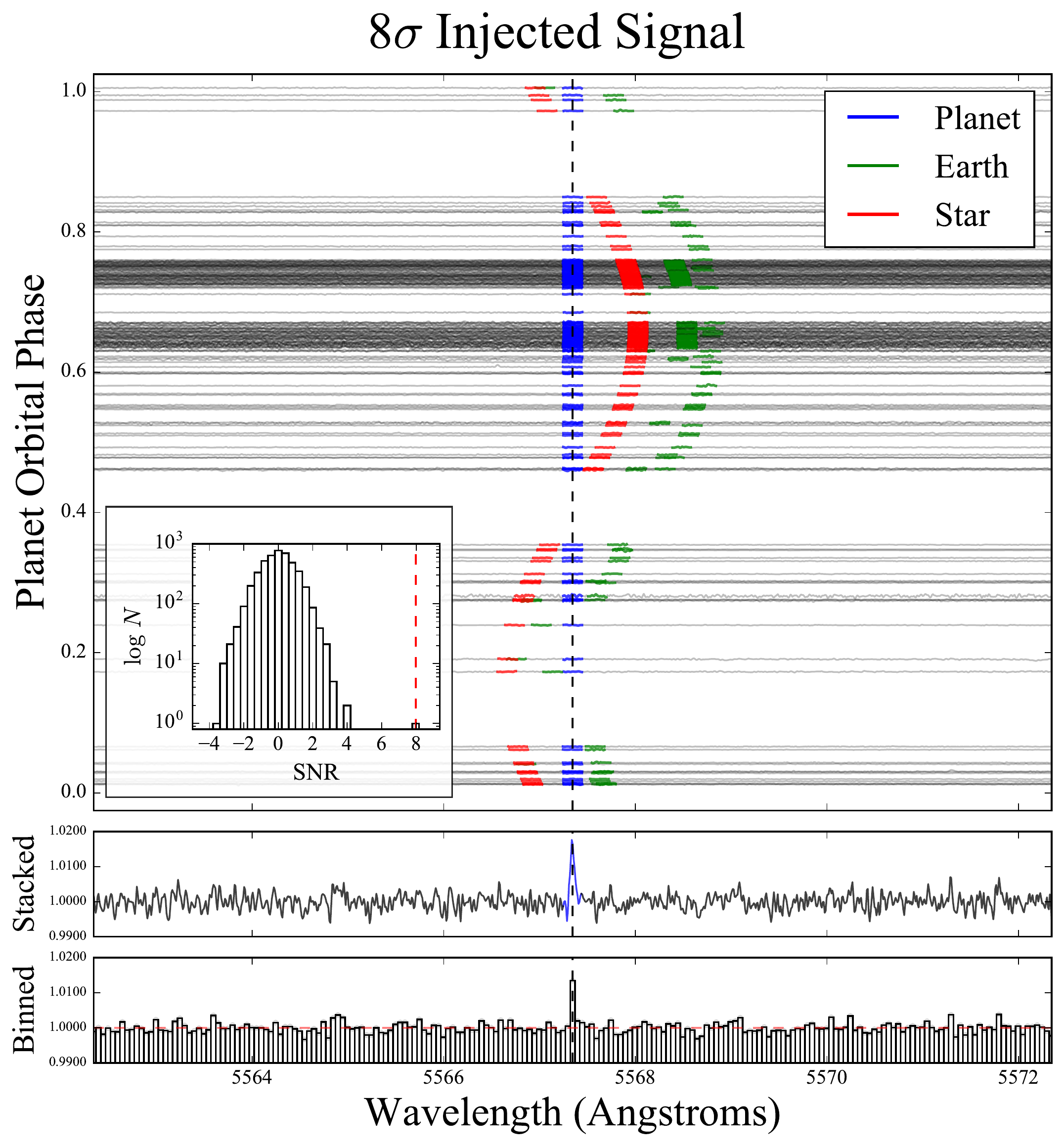}
\caption{Similar to Fig.~\ref{fig:strongest_river}, but for an emission feature injected into the raw data at 5567.345\AA\ (10\AA\ blueward of the OI line, where no emission is expected) with contrast $1.8\times 10^{-2}$, corresponding to a power of $2.6\times 10^3$ TW. Our method recovers the signal in the stacked, binned spectrum with SNR ${\sim} 8$ and a detection significance of $8\sigma$, our nominal detection threshold. The non-detection in Fig.~\ref{fig:strongest_river} therefore constrains the auroral power on Proxima Cen b to be $\lesssim 3\times 10^{3}$~TW, consistent with the calculations in \S\ref{sec:signal}.\\[0in]}
\label{fig:injection_river}
\end{figure}

The stacked flux in the planetary frame is shown below the main plot, where the peak at 5577\AA\ is visible. Below it, we show the stacked flux binned to 0.05\AA\ bins; the feature also stands out here. The inset at the bottom left of the main subplot shows a histogram of the SNR of all the bins in a 250\AA\ window centered on the OI line; the 5577\AA\ feature (indicated by a dashed red line) is one of only two with SNR ${\sim} 4$. However, this signal is consistent with correlated stellar noise and is in no way a detection of planetary 5577\AA\ emission. To show this, we performed the same grid search used to generate Fig.~\ref{fig:triangle} in each of 2,250 wavelength bins on either side of 5577\AA\ (where we do not expect significant planetary emission) and computed the strongest recovered signal in each wavelength bin. The results are shown in Fig.~\ref{fig:max_signal_vs_wavelength}, where the 5577\AA\ bin is indicated by a red circle. The dashed red line indicates the fractional strength of the signal in that bin, ${\sim} 0.006$, which roughly corresponds to a planet-star contrast of the same magnitude if the signal were real. However, it is clear from the figure that over 20\% of the bins in the range $5465-5670$\AA\ have a stronger peak signal than that in the 5577\AA\ bin, from which we estimate a detection significance of ${\sim} 0.7\sigma$ and a false alarm probability (FAP) for our recovered peak of ${\sim} 0.2$. These data are also shown as a histogram in Fig.~\ref{fig:fap}, where the number of signals are plotted as a function of their fractional strength. It is evident from these figures that the peak shown in Fig.~\ref{fig:strongest_river} is fully consistent with noise.

A striking feature of Fig.~\ref{fig:max_signal_vs_wavelength} is the correlated nature of the noise; for instance, no bins in the range $5500 - 5575$\AA\ have particularly strong signals, while the 5577\AA\ peak is one of many in its immediate vicinity. This is likely a sign of correlated stellar noise that was improperly removed by PCA, either due to high temporal stellar variability or nonlinear correlations across the spectrum that cannot be captured by PCA. At present, this noise is the limiting factor in our ability to recover auroral emission by Proxima Cen b. Future coronagraphic observations of Proxima Centauri (\S\ref{sec:detect}) should greatly reduce this noise by nulling most of the starlight.

In order to quantify the constraints our non-detection imposes on the properties of Proxima Cen b, we inject Gaussian OI emission signals with FWHM 0.05\AA\ and of varying contrast into each of the raw spectra and attempt to recover them via the procedure described above. We find that an OI auroral signal with planet-star contrast $1.8\times 10^{-2}$ yields a detection with SNR ${\sim} 8$ (Fig.~\ref{fig:injection_river}). By performing the same wavelength search as before (Figs.~\ref{fig:max_signal_vs_wavelength} and \ref{fig:fap}), we estimate the significance of this detection to be $8\sigma$, which we conservatively choose to be our nominal detection threshold for the HARPS dataset. From Table~\ref{tab:OI_detect}, this contrast is equivalent to an OI auroral power of ${\sim} 3\times 10^{3}$~TW, our empirical upper limit on the strength of Proxima Cen b's 5577\AA\ emission. Note that by scaling the estimates in that table, it should take only a few hours (versus the 50 hours used in the injection/recovery step) to detect such a signal with HARPS. This order-of-magnitude difference is likely related to the correlated stellar noise discussed above. Integration times for telescopes without coronagraphs may therefore be significantly longer than the estimates quoted in that table.

Finally, we also performed similar searches for the red oxygen lines (6300.308 and 6363.790\AA) and the 3914.4\AA\ UV nitrogen line, which are prominent in Earth's aurora, but find no significant peaks. Given the lower power in the red lines relative to the green line, and the low transmissivity of Earth's atmosphere and lower detector efficiencies in the UV, this non-detection is consistent with the non-detection of the 5577\AA\ feature.
  
\section{Discussion \& Conclusions}
\label{sec:disc}
Our calculations above assume that Proxima Cen b is a terrestrial planet with an Earth-like atmosphere. Although its radius is not known --- making an estimate of its density impossible --- the planet is statistically likely to be rocky. The \emph{a priori} probability distribution for the inclination of an exoplanet is $\mathrm{P}(i)\mathrm{d}i = \sin(i)\mathrm{d}i$. With 68\% confidence, the inclination of Proxima Cen b is greater than 47$^\circ$, and with 95\% confidence it is greater than 18$^\circ$. Given $m_p\sin i = 1.27\mathrm{M}_\oplus$, this corresponds to true planet masses smaller than 1.7M$_\oplus$ (68\% confidence) and 4.1M$_\oplus$ (95\% confidence). Recent exoplanet population studies suggest that the transition between rocky and gaseous exoplanets occurs at a radius of about ${\sim}1.6\mathrm{R}_\oplus$ \citep{Rogers2015, Wolfgang2015}. However, a corresponding value for the transition \emph{mass} is still uncertain, and since the radius of Proxima Cen b has not been measured, we cannot argue for its terrestrial nature based on its mass alone. Nevertheless, we can obtain predictions for its radius under certain assumptions. Assuming it is a rocky planet with Earth-like composition, we may use the scaling law from \citet{Fortney2007} to obtain radii of 1.16R$_\oplus$ ($m_p = 1.7\mathrm{M}_\oplus$) and 1.45R$_\oplus$ ($m_p = 4.1\mathrm{M}_\oplus$).
Using the mass-radius grids of \citet{Lopez2012} and assuming instead that Proxima Cen b is a super-Earth/mini-Neptune with a thin H/He envelope with mass equal to 1\% the planet mass, the radii jump to 1.7R$_\oplus$ ($m_p = 1.7\mathrm{M}_\oplus$) and 2.1R$_\oplus$ ($m_p = 4.1\mathrm{M}_\oplus$).
Planet occurrence rate calculations for cool M dwarfs \citep{Dressing2013} suggest that there is a steep drop in the number of short-period planets per star with radii above 1.4R$_\oplus$ ($0.19^{+0.07}_{-0.05}$), compared to those with radii below 1.4R$_\oplus$, which are more than twice as common ($0.46^{+0.09}_{-0.06}$). This suggests that Proxima Cen b is more likely to be terrestrial than Neptune-like. A thin ($\lesssim 1\%$ by mass) H/He veneer is still possible, but given the extended pre-main sequence phase of the host star, past hydrodynamic escape is likely to have blown it off \citep{Luger2015, Barnes2016}.

Nevertheless, we cannot definitively rule out the possibility that Proxima Cen b has an atmosphere dominated by H/He, in which case we would not expect OI auroral emission. A search for Lyman-Werner H$_2$ emission in the UV would be more appropriate in this case. Although broader than the lines we consider here, this emission is likely stronger, and is unlikely to be confused with stellar emission, given that it is molecular in origin. But perhaps more importantly, a robust \textit{non}-detection of this and other H/He features could rule out a large gaseous envelope and confirm the terrestrial nature of the planet. That said, we are currently unable to efficiently probe near-face-on orbits due to the much smaller Doppler shift of the planetary lines. Observations made exclusively at quadrature, when the planet RV is highest, may help with this in the future.

Alternatively, Proxima Cen b could be terrestrial but be significantly larger than the Earth, with mass as high as ${\sim}4\mathrm{M}_\oplus$ and radius ${\sim}1.5\mathrm{R}_\oplus$.
Since the scaling methods used in \S\ref{sec:signal} implicitly assume Proxima Centauri b is similar to Earth in size, the auroral strength could be different than what we estimate. Assuming a global field, a larger planetary radius (and therefore core radius) could increase the magnetospheric cross-section to the stellar wind, leading to an increase in the emitted power. Furthermore, assuming an Earth-like atmospheric composition, the higher surface gravity would decrease the ionospheric scale height, which could lead to larger magnetic field parallel potential drops in auroral acceleration regions. This would increase the upward flowing current and therefore the downward accelerated electron beams into the upper atmosphere, which could also enhance the auroral signal. Moreover, an increased atmospheric density at the depth where precipitating electrons deposit their energy could also change recombination rates and alter the energy level distribution of the O atoms, which would in turn affect the auroral strength in different lines. A quantitative estimate of these effects is beyond the scope of this study, as it would require both modeling the changes to the atmospheric structure and solving the Boltzmann kinetic transport equation.

Assuming Proxima Cen b is terrestrial, our HARPS search constrains its auroral power to be $< 3\times 10^{3}$~TW. This is consistent with the calculations in \S\ref{sec:signal}, which suggest the OI auroral power on Proxima Cen b is likely ${\sim} 0.1$~TW, or ${\sim} 100\times$ that of the Earth during steady-state solar wind conditions. Those calculations, however, ignore transient increases in stellar magnetic activity, which can enhance the auroral signal and the diffuse airglow emission of the planet. As discussed in \S\ref{sec:signal}, transient magnetospheric activity could result in auroral power for the 5577\AA\ line up to $10-100$~TW, lasting from $10-10^3$ minutes (\S\ref{sec:signal_m2}).  In addition, for a planet that is under near constant CME activity, it is possible that the storm conditions last for weeks or longer \citep{Gonzalez1994,Gonzalez1999}. Spectra taken during periods of vigorous stellar activity could thus enhance the chances of detecting auroral emission.

Even if Proxima Cen b is terrestrial, an auroral signal is not guaranteed to be present. The existence of an atmosphere is still an open question, owing to vigorous past hydrodynamic escape \citep{LugerBarnes2015, Barnes2016}, observed persistent stellar activity \citep{Davenport2016} and an observationally unconstrained planetary magnetic field. If an atmosphere is in fact present, it may not be Earth-like; instead, it could be dominated by CO$_2$ \citep[e.g.][]{Meadows2016}. Airglow and auroral 5577\AA\ emission are still expected for such an atmosphere, since atomic oxygen is produced by photodissociation of CO$_2$; in fact, OI 5577\AA\ emission has been observed at both Mars and Venus, both of which are CO$_2$-dominated \citep[e.g.][]{Bertaux2005, Slanger2001}. In particular, \citet{Slanger2001} and \citet{Slanger2006} found that the Venusian airglow strength is comparable to that of Earth. Given that Venus receives about twice the solar flux Earth receives, airglow and/or auroral emission from a CO$_2$-rich Proxima Cen b could be a factor of ${\sim}2$ weaker than the values we predict in this paper, although detailed photochemical modeling is required to accurately model this scenario.

In the case that the atmosphere is oxygen-rich, nitrogen may need to be present to enhance the auroral signal. On Earth, the OI green line emission results primarily from O$_2^+$ dissociative recombination, as well as collisions with excited N$_2$ and direct electron impact \citep{Strickland2000}. It is unclear whether or not other molecular species could play a similar role if N$_2$ is not the bulk atmospheric constituent. However, the detection of the 3914\AA\ N$_2^+$ band could be a good diagnostic in the UV, where the star is even fainter. If we assume that the power of the 3914\AA\ nitrogen band is comparable to that of the OI line (which is typical for higher energy magnetospheric particle populations), then the planet-star contrast in the N$_2^+$ band would be an order of magnitude greater than at 5577\AA. Since the strength of the N$_2^+$ band scales with magnetospheric parameters, stellar activity could cause strong transient features in the UV, which may be observable. Note, however, that limitations in UV detector efficiencies may complicate the detection of nitrogen and other UV aurorae.

Our integration times for the predicted steady-state 5577\AA\ OI auroral line render its detection infeasible for current facilities. However, if key design goals are met for future coronagraphs, steady-state aurorae may be more easily detected. As shown in Fig.~\ref{fig:contrast}, achieving the optimal star-planet contrast ratio at the emission line requires that the width of a spectral element (resolving power) is smaller (greater) than the line's equivalent width. For our predicted steady-state auroral emission (${\sim}0.1$~TW), this requires future spectrographs to achieve $R \gtrsim 10^5$. High-resolution spectroscopy is also needed to resolve the Doppler shift of the planetary auroral emission (${\sim}1$\AA) and place strong constraints on the eccentricity and mass/inclination of the planet. Such constraints would lead to greater confidence in the terrestrial nature of Proxima Cen b. Furthermore, since read noise and dark current dominate the coronagraph instrumental noise budget, the development of low-noise detectors, e.g. MKIDS \citep{Mazin2012,Mazin2015}, would significantly help the detection sensitivity and would allow such high-resolution spectroscopy to be downbinned to the lower resolution typically considered for direct exoplanet spectroscopy. For instance, if future detectors render read noise and dark current negligible, a LUVOIR concept telescope could observe a $0.1$~TW OI $5577$\AA\ auroral emission feature in 100 hours as opposed to the $2 \times 10^3$ hours for the noised observations considered in \S\ref{sec:detect}. In addition, low-noise detectors would make stacking short observations (required to mitigate broadening of the line due to the planet's orbital motion) more feasible. If TMT is built with a coronagraph that can achieve a design contrast of $10^{-7}$ and negligible instrumental noise, it could observe steady-state auroral emission (${\sim}0.1$~TW) in a few nights. However, such an observation would require the development of an effective AO system in the optical.

Alternatively, observations made during periods of vigorous stellar activity may enhance the detectability of exo-aurorae on Proxima Cen b.  Transient magnetospheric activity could increase auroral power to $1-100$~TW, depending on the planetary magenetic dipole strength, allowing future coronagraph-equipped TMT and LUVOIR telescopes to detect auroral emission in 1 hour or less. This is comparable to estimated CME timescales \citep{Khodachenko2007} and much shorter than the timescales of long-lasting solar storm conditions \citep{Gonzalez1994,Gonzalez1999}. Future observing missions similar to the MOST campaign \citep[e.g.][]{Davenport2016} could be used to characterize and monitor Proxima Centauri's activity levels to constrain the star's activity cycles. Such missions could aid in scheduling spectroscopic observations of Proxima Centauri. Observing the star following a CME-like event or long duration fast solar streams could enhance detectability of the planetary auroral signal.  

The methods of exo-auroral detection discussed here are not limited to Proxima Cen b, but may be applicable to any exoplanet orbiting a nearby late-type star or brown dwarf.  For example, the recently discovered TRAPPIST-1 system \citep{Gillon2016} consists of three planets orbiting an active late M8 ultracool dwarf only 12 pc away; one planet in the system, TRAPPIST-1d, potentially lies in the habitable zone. Since TRAPPIST-1 is a later type star than Proxima Centauri, it is likely more active \citep[e.g.,][]{West2008} and hence could generate larger particle fluxes and a stronger interplanetary magnetic field than Proxima Centauri, leading to more powerful aurorae on its planets. Additionally, the redder blackbody spectrum of TRAPPIST-1 makes it a factor of about 6 dimmer than Proxima Centauri at the OI $5577$\AA\ line, resulting in far more favorable contrast ratios. However, due to its distance, auroral emission from this system will be ${\sim} 100\times$ dimmer than that from Proxima Centauri, likely making its detection infeasible. For coronagraphic observations, the distance to the TRAPPIST system would require an inner working angle smaller than the diffraction limit to extend as long as 5577\AA\ for all known TRAPPIST-1 planets observed with a 10m class telescope, further complicating the observation.

Another planet to consider is GJ1132b, which orbits a M3.5 star 12 pc away \citep{Berta2015}. Since it receives ${\sim} 19\times$ the Earth's flux and may have an O$_2$ rich atmosphere \citep{Schaefer2016}, it could display strong auroral emission. However, as with the TRAPPIST-1 system, its distance makes auroral characterization difficult. Moreover, the earlier type host emits a larger fraction of its light in the optical, resulting in a poorer auroral contrast ratio.

Finally, exoplanets orbiting nearby brown dwarfs may be prime targets for exo-auroral searches. Early-type brown dwarfs display significant magnetic activity \citep{West2008} and are significantly fainter than M dwarfs in the optical, potentially enhancing the detectability of the 5577\AA\ signal from planets in orbit around them. Although no short-period exoplanets are currently known to orbit nearby brown dwarfs \citep{He2016}, the methods described in this paper may be used as means of exoplanet detection, as suggested by \citet{SparksFord2002}. Since the stack-and-search method described in \S\ref{sec:search} does not require previous RV observations of a system, a long baseline of spectroscopic observations of nearby M dwarfs and brown dwarfs could be used to search for Doppler-shifted 5577\AA\ OI emission. Our method is particularly sensitive to short-period terrestrial planets, whose auroral power (if an atmosphere is present) is large and whose large RV will Doppler-shift the signal by one or more \AA. However, since the stack-and-search method is best suited to detect steady-state emission, exoauroral searches will likely have to wait for a future generation of space-based telescopes or noiseless ground-based ELTs capable of detecting sub-TW aurorae. These searches may someday reveal the presence of unknown nearby terrestrial exoplanets, including ones in the habitable zone.

\clearpage

All code used to generate the tables and figures in this paper is open source and available at \url{https://github.com/rodluger/exoaurora}. A static version of the code is archived at \url{https://doi.org/10.5281/zenodo.192459}.\\[0in]

 %% Acknowledgements %%
\acknowledgments

We thank G. Anglada-Escude and the Pale Red Dot team for making their data publicly available, Giada Arney for useful discussions, R. W. Service for poetic inspiration, and the anonymous referee for their excellent comments and suggestions. DPF is supported by an NSF IGERT DGE-1258485 fellowship. This work was supported by the NASA Astrobiology Institute's Virtual Planetary Laboratory under Cooperative Agreement number NNA13AA93A and was based on data products from observations made with ESO Telescopes at the La Silla Paranal Observatory under programme IDs 072.C-0488(E), 082.C-0718(B), 096.C-0082(A), 096.C-0082(B), 096.C-0082(C), 096.C-0082(D), 096.C-0082(E), 096.C-0082(F), 183.C-0437(A), and 191.C-0505(A). This work made use of the advanced computational, storage, and networking infrastructure provided by the Hyak supercomputer system at the University of Washington. Finally, this work also made use of the Python coronagraph noise model, developed by J. Lustig-Yaeger and available at \url{https://github.com/jlustigy/coronagraph/}

%% Bibliograph %%
\bibliography{bib}

%% End Doc
\end{document}